\documentclass[twocolumn,aps,tightenlines,superscriptaddress,amsmath,amssymb,amsfonts,noeprint,longbibliography,prb]{revtex4-2}
\usepackage{graphicx}                   
\usepackage{dcolumn}                    
\usepackage{bm}                         
\usepackage{float}
\usepackage{hyperref}                   
\usepackage[section]{placeins}
\usepackage{epstopdf}
\usepackage[table, dvipsnames]{xcolor}  
\usepackage{newtxtext,newtxmath}        

\newcommand{\ie}{\textit{i}.\textit{e}.}
\newcommand{\eg}{\textit{e}.\textit{g}.}

\DeclareMathOperator{\sech}{sech}

\begin{document}

\title{Low-energy effective theory of localization-delocalization transition in noninteger-charged electron wave packets}
\author{Y. Yin}
\email{yin80@scu.edu.cn}
\thanks{Author to whom correspondence should be addressed}
\affiliation{Department of Physics, Sichuan University, Chengdu, Sichuan, 610065, China}
\date{\today}
\begin{abstract}
  We present a low-energy effective theory to describe the localization-delocalization transition, which occurs for wave functions of electrons and holes injected individually by a voltage pulse with noninteger flux quantum. We find that the transition can be described by an effective scattering matrix in a truncated low-energy space, which is composed of two parts. The first part describes the infrared-divergence of the scattering matrix, while the second part represents the high-energy correlation. For short-tailed pulses which decay faster than Lorentzian, the scattering matrix exhibits solely an inverse linear divergence in the infrared limit. The divergence is responsible for the dynamical orthogonality catastrophe, which leads to electron-hole pairs with delocalized wave functions. In contrast, the high-energy correlation can be approximated by a constant term, which leads to electron-hole pairs with localized wave functions. Due to the competition between the two terms, the wave functions can undergo a localization-delocalization transition, which occurs for electrons and holes injected individually by the voltage pulse. As a consequence, the localization-delocalization transitions for all short-tailed pulses can be described by the same effective scattering matrix, suggesting that they belong to the same universality class. For pulses with longer tails, the scattering matrix can exhibit additional infrared-divergences. We show that a Lorentzian pulse gives rise to a logarithmic divergence, while a fractional-powered Lorentzian pulse gives rise to a power-law divergence. The additional divergence can lead to localization-delocalization transitions belonging to different universality classes. These results demonstrate the fine-tuning capabilities of the localization-delocalization transition in time-dependent quantum transport.
\end{abstract}
\maketitle

\section{Introduction\label{sec1}}

In a seminal work in 1967 \cite{Anderson1967}, Anderson shows that the ground states of a non-interacting Fermi gas with and without a localized potential are orthogonal in the thermodynamics limit. As a consequence, a sudden switching of the potential can shake up the whole Fermi gas, leading to the excitation of a diverging number of electron-hole (eh) pairs \cite{NOZI_RES_1969, Schotte_1969, Ohtaka_1990, Mahan_2000}. This is known as the orthogonality catastrophe \cite{Mahan_2000}, which is first addressed in the X-ray absorption of simple metals \cite{Ohtaka_1990}. It is also found in the optical spectra of semiconductor nanostructures \cite{Lee_1987, Skolnick_1987, Calleja_1991, Hentschel_2005, Hentschel_2007, Heyl_2012}, matter-wave interferometry in ultracold fermions \cite{Goold_2011, Knap_2012, Sindona_2013, Cetina_2016, Schmidt_2018}, and resonant tunneling through localized levels \cite{Matveev_1992, Geim_1994, Cobden_1995, Benedict_1998, Hapke_Wurst_2000, Frahm_2006, Ubbelohde_2012}. The potential is essentially static in these cases. It is induced by point-like impurities and switched on by the excitation of the impurity states.

In contrast to the static case, the potential can also be dynamic. This typically occurs in the study of time-dependent quantum transport, where the potential is induced by a single voltage pulse \cite{keeling_2006_minim, dubois_2013_minim, gabelli_2013_shapin, Gaury_2014, glattli_2016_levit, baeuerle_2018_coher}. The orthogonality catastrophe in this case --- usually referred to as dynamical orthogonality catastrophe --- only occurs when the pulse carries noninteger multiples of a flux quantum \cite{ivanov_1995_coher, levitov_1996_elect, dubois_2013_integ, hofer_2014_mach, glattli_2017_pseud}. Hence it can be suppressed in a controllable manner by varying the flux of the pulse. The wave functions of electrons and holes can also be reconstructed by quantum state tomography \cite{jullien_2014_quant, bisognin_2019_quant}. This provides an alternative way to study the problem of the orthogonality catastrophe. In a previous work \cite{Yin_2025}, we have shown that eh pairs with delocalized wave functions can be injected when the dynamical orthogonality catastrophe occurs, and vanish when the dynamical orthogonality catastrophe is suppressed. In the meantime, the wave functions of electrons and holes injected individually by the voltage pulse evolve from a delocalized one to a localized one. This leads to a localization-delocalization (LD) transition, which can be described by a single-parameter scaling theory. The scaling function and correlation length of the LD transition can exhibit different behaviors for pulses with different profiles. This suggests that the LD transition can belong to different university class for different pulses, which has not been full clarified in the previous work.

In this paper, we present a low-energy effective theory, which can provide a better understanding of the LD transition in this case. We show that the quantum states of the injected electrons, holes and eh pairs can be described by an effective scattering matrix in a restricted low-energy space. It can be separated into two parts. The first part describes the infrared (IR) divergence of the scattering matrix, while the second part represents the high-energy correlation. The eh pairs with delocalized wave functions are determined by the first part, while the eh pairs with localized wave functions are dominated by the second part. Both parts are necessary to describe the wave functions of individually-injected electrons and holes, which can undergo the LD transition. As far as the LD transition is concerned, the high-energy correlation can be well approximated by a constant. In contrast, the IR divergence of the scattering matrix has to be considered exactly. 

The scattering matrix can exhibit two kinds of IR divergence: (a) an inverse linear divergence. The divergence is universal in the sense that its coefficient is independent of the detailed profile of the voltage pulse, but only decided by its Faraday flux. The divergence is also responsible for the dynamical orthogonality catastrophe, which occurs when the flux is equal to noninteger multiples of a flux quantum. (b) an additional divergence, which appears when the pulse decays slower than Lorentzian at long times. The divergence can be fine tuned by changing the profile of the pulse. As typical examples, we show that a Lorentzian pulse gives rise to a logarithmic divergence, while a fractional-powered Lorentzian pulse gives rise to a power-law divergence. By properly account for the IR divergence, the effective scattering matrix can reproduce the scaling function and correlation length of the LD transition for both the short-tailed (a) and long-tailed (b) pulses. As a consequence, the LD transition can be described by the effective scattering matrix of the same form for all short-tailed pulses. This indicates that they belong to the same university class. In contrast, the long-tailed pulses can be described by the effective scattering matrix with different forms, leading to LD transitions belonging to different university classes. This suggests that the LD transition in such system can be fine tuned.

The paper is organized as follows. In Sec.~\ref{sec2}, we review our model for the charge injection and show how to extract the wave functions of electrons and holes from the scattering matrix. In Sec.~\ref{sec3}, we present the low-energy effective theory for the short-tailed pulses. We show that the theory can reproduce both the scaling function and correlation length for the LD transition in this case. In Sec~\ref{sec4}, we generalize the low-energy effective theory to the long-tailed pulses. We summarize in Sec.~\ref{sec5}.

\section{Model and formalism\label{sec2}}

The model we used here is the same as that used in our previous works \cite{Yue_2019, Yin_2025}. So we only briefly outline the main points. We consider the charge injection from a reservoir into a single-mode quantum conductor, which is driven by a time-dependent voltage $V(t)$ applied on the electrode. We choose the driving voltage $V(t)$ of the form
\begin{eqnarray}
  V(t) = V_p(t_0+t) - V_p(t_0-t).
  \label{s2:eq10}
\end{eqnarray}
It corresponds to two successive pulses with the same shape but opposite signs, which are separated by a time interval $2t_0$. For simplicity, we assume the pulse is symmetric, \ie, $V_p(t) = V_p(-t)$. We characterize the width of each pulse by the half width at half maximum $W$. The strength of the pulse can be described by the flux $\varphi = (e/h) \int^{+\infty}_{-\infty} V_p(t) dt$, which is the Faraday flux of the voltage pulse normalized to the flux quantum $h/2e$ with $h$ being the Planck constant and $e$ being the electron charge. In the following part of the
paper, we choose $e=h=W=1$.

The electron injection in this system can be fully characterized by the scattering matrix
\begin{eqnarray}
  S(E) = \int^{+\infty}_{-\infty} dt e^{2\pi i E t - i\phi(t)},
  \label{s2:eq20}
\end{eqnarray}
with $\phi(t) = 2\pi \int^t_{-\infty} V(\tau) d\tau$ being the scattering phase due to the driving pulses. The many-body state of the injected electrons can be expressed as
\begin{equation}
  | \Psi \rangle = \prod_{k=1, 2, 3, \dots} \Big[ \sqrt{1 - p_k} + i \sqrt{p_k} B^{\dagger}_e(k) B^{\dagger}_h(k) \Big] | F \rangle, \label{s2:eq30}
\end{equation}
where $|F\rangle$ represents the Fermi sea and $B^{\dagger}_e(k)$[$B^{\dagger}_h(k)$] represents the creation operator for the electron[hole] component of the eh pairs. They can be expressed as
\begin{eqnarray}
  B^{\dagger}_e(k) & = & \int^{+\infty}_0 dE \psi^e_k(E) a^{\dagger}(E), \\
  B^{\dagger}_h(k) & = & \int^0_{-\infty} dE \psi^h_k(E) a(E).
                         \label{s2:eq40}
\end{eqnarray}

The wave function $\psi^{e/h}_k(E)$ and excitation probability $p_k$ can be obtained from the polar decomposition of the scattering matrix. This can be done by solving the following equation for $E>0$:
\begin{equation}
  \int^{+\infty}_{0} dE' S(E+E') \psi^{\ast}_{k}(E') = i \sigma \sqrt{p_k} \psi_k(E),
  \label{s2:eq50}
\end{equation}
with $\sigma = \pm 1$. The wave function of the electron $\psi^e_k(E)$ and hole $\psi^h_k(E)$ can be obtained as $\psi^e_k(E) = \sigma \psi^h_k(-E) = \psi_k(E)$. In the numerical calculation, Eq.~(\ref{s2:eq50}) can be solved in the energy domain by using the singular value decomposition.

In our previous work, we have shown that eh pairs with delocalized wave functions can be injected when the voltage pulse carries a noninteger flux quantum \cite{Yin_2025}. In the following section, we shall show that the delocalized wave functions are attributed to the low-energy divergence of the scattering matrix, which is responsible for the dynamical orthogonality catastrophe. The divergence is critical for the description of the LD transition.

\section{Low-energy scattering theory for short-tailed pulses \label{sec3}}

To show so, let us first decompose the scattering matrix $S(E)$ into two parts, which correspond to the positive and negative pulses, respectively. When $t_0 \gg 1$, the integral in Eq.~(\ref{s2:eq20}) can be well approximated as
\begin{eqnarray}
  S(E) & = & 2\pi \delta(E) + \int^{0}_{-\infty} dt e^{2\pi i E t} [ e^{ - i \phi_p(t_0+t)} - 1 ] \nonumber\\
  && {}+ \int^{+\infty}_{0} dt e^{ 2\pi i E t} [ e^{ - i \phi_p(t_0-t)} - 1 ],
  \label{s3:eq10}
\end{eqnarray}
where $\phi_p(t) = 2\pi  \int^t_{-\infty} V_p(\tau) d\tau$ represents the scattering phase of a single pulse. The scattering matrix can be written in a more compact form:
\begin{equation}
  S(E) = 2\pi \delta(E) + e^{-i E t_0} S_p(E) + e^{i E t_0} S_p(-E).
  \label{s3:eq20}
\end{equation}
The Dirac delta function represents the zero energy component, while $S_p(E)$ represents the scattering matrix of a single pulse for $E \ne 0$. It can be expressed as
\begin{equation}
  S_p(E) = \int^{t_0}_{-\infty} dt e^{-i E t} [ e^{ - 2\pi i \phi_p(t)} - 1 ].
  \label{s3:eq30}
\end{equation}

The second and third terms in Eq.~(\ref{s3:eq20}) then correspond to the positive and negative pulses around $t = \pm t_0$, respectively. As $t_0$ goes to infinity, one may expect that their contributions are decoupled due to the high-frequency oscillation from $e^{ \pm i E t_0}$. However, this is not always the case because $S_p(E)$ contains a term proportional to $1/E$, which diverges in the zero energy limit. This term can be singled out by using integration by parts from Eq.~(\ref{s3:eq30}), which gives
\begin{eqnarray}
  S_p(E) & = & (e^{-2\pi i \varphi} - 1) \frac{e^{i E t_0} - 1}{i E} \nonumber\\
         &&{}+ \int^{+\infty}_{-\infty} dt 2 \pi V_p(t) e^{-i\phi_p(t)} \frac{e^{i E t} - 1}{E}.
  \label{s3:eq40}
\end{eqnarray}
Note that we have set the upper bound of the integral to $+\infty$, which is a good approximation in the large $t_0$ limit.

In this limit, the first term in the right-hand side of Eq.~(\ref{s3:eq40}) exhibits an IR divergence proportional to $1/E$ and dominates the scattering matrix $S_p(E)$ at low energies. Its coefficient is only decided by the flux of the voltage pulse $\varphi$. In contrast, the second term depends on the detailed profile of the pulse $V_p(t)$. In particular, this term vanishes in the limit when the pulse becomes a delta function. In this section, we focus on the case when the second term can be expanded into a convergent Taylor series:
\begin{equation}
  \int^{+\infty}_{-\infty} dt 2 \pi V_p(t) e^{-i\phi_p(t)} \frac{e^{i E t} - 1}{E} = e^{-i\pi \varphi} \sum_{n=0, 1, 2, ...}  \frac{b_n}{n!} E^n,
  \label{s3:eq50}
\end{equation}
with
\begin{equation}
  b_n = e^{i\pi \varphi} \frac{i^{n+1}}{(n+1)!} \int^{+\infty}_{-\infty} dt V_p(t) t^n e^{-i\phi_p(t)}.
  \label{s3:eq60}
\end{equation}
This occurs when the voltage pulse decays sufficiently fast so that the integral in Eq.~(\ref{s3:eq60}) converges. Typical examples are Gaussian or hyperbolic secant pulses, which has been discussed in the previous work \cite{Yin_2025}. Note that the coefficients $b_n$ are all real, as we have assumed that $V_p(t)$ is symmetric.

The expansion given in Eqs.~(\ref{s3:eq40}), (\ref{s3:eq50}) and (\ref{s3:eq60}) provides a natural starting point to construct the low-energy effective theory. It suggests that the low energy scattering matrix can be obtained by truncating the Taylor expansion at a given order. In particular, the first-order truncation gives the effective scattering matrix
\begin{equation}
  e^{-i \pi \varphi} S^{\rm eff}_p(E) = C_0 \frac{e^{i E t_0} - 1}{E} + B_0.
  \label{s3:eq70}
\end{equation}
In doing so, the low-energy divergence is incorporated exactly by the first term in the right-hand side of Eq.~(\ref{s3:eq70}). The high-energy correlation is described approximately by the constant term $B_0$. We assume the coupling parameters $C_0$ and $B_0$ are all real.

The coupling parameters can be obtained by the perturbative matching, which decides the parameters by fitting a given physical quantity order by order in terms of a small ratio of low-to-high energy scales \cite{Lepage_1997, Delamotte_2004}. In this paper, we choose the electron number $N(E)$ as the physical quantity. For the effective scattering matrix given in Eq.~(\ref{s3:eq70}), it is defined as
\begin{equation}
  N(E) = \int^{\Lambda}_{E} dE' f(E'),
  \label{s3:eq80}
\end{equation}
where $\Lambda$ represents an ultraviolet cutoff and $f(E)$ represents the electron distribution function. The distribution function can be calculated from the effective scattering matrix as
\begin{equation}
  f(E) = \int^{\Lambda}_{0} \frac{dE'}{2\pi} \left| S^{\rm eff}_p(E+E') \right|^2.
  \label{s3:eq90}
\end{equation}
The exact $N(E)$ can be obtained by replacing $S^{\rm eff}_p(E)$ to the exact scattering matrix $S_p(E)$ given in Eq.~(\ref{s3:eq40}) and sending the upper bound of integrals in Eqs.~(\ref{s3:eq80}) and (\ref{s3:eq90}) to $+\infty$.


We tune the coupling parameters so that $N(E)$ calculated from the effective scattering matrix best reproduces the exact $N(E)$ in the limit $E \to 0$ and $t_0 \to +\infty$. To achieve this, we introduce a small positive imaginary part to the energy, \ie, let $E \to E + i \eta$. We first let $t_0 \to +\infty$ then let $\eta \to 0$. In this limit, the electron number $N(E)$ becomes independent on $t_0$. For the effective scattering matrix, $N(E)$ can be expanded in the series of $E$ as
\begin{equation}
  N(E) = -\frac{C^2_0}{2\pi} \ln(E) + \frac{C_0 B_0}{\pi} E \ln(E) + N_0 + O(E),
  \label{s3:eq100}  
\end{equation}
with $N_0 = (C^2_0/2\pi) \ln(\Lambda/2) + (B^2_0/2\pi) \Lambda^2$. In the limit $E \to 0$, the leading-order contribution comes from the first term in the right-hand side of Eq.~(\ref{s3:eq100}), which exhibits a logarithmic divergence. By comparing the coefficient of this term to the exact result [See the Appendix for details], one obtains $C_0 = 2 \sin(\pi \varphi)$. The next-to-leading-order contribution comes from the second term, whose derivative diverges as $E \to 0$. By comparing the coefficient of this term to the exact result, one has
\begin{equation}
  B_0 = b_0 = e^{i\pi \varphi} i \int^{+\infty}_{-\infty} dt V_p(t) e^{-i\phi_p(t)}.
  \label{s3:eq105}
\end{equation}
Note that the two parameters $C_0$ and $B_0$ are independent on the ultraviolet cutoff $\Lambda$. The rest part of $N(E)$ is analytic in terms of $E$, which can be approximated by the constant term $N_0$ in the limit $E \to 0$. By requiring this term is equal to its counterpart from the exact solution, one obtains the optimized value of the cutoff $\Lambda$.

Given the effective scattering matrix and ultraviolet cutoff $\Lambda$, the wave function and excitation probability from the effective theory can be calculated in analogy to Eq.~(\ref{s2:eq50}):
\begin{equation}
  \int^{\Lambda}_{0} dE' S^{\rm eff}(E+E') \psi^{\ast}_{k}(E') = i \sigma \sqrt{p_k} \psi_k(E),
  \label{s3:eq110}
\end{equation}
where $S^{\rm eff}(E)$ represents the effective scattering matrix corresponding to the two pulses. From
Eq.~(\ref{s3:eq20}), it can be related to the effective scattering matrix of the single pulse $S^{\rm eff}_p(E)$
[Eq.~(\ref{s3:eq70})] as
\begin{equation}
  S^{\rm eff}(E) = e^{-i E t_0} S^{\rm eff}_p(E) + e^{i E t_0} S^{\rm eff}_p(-E).
  \label{s3:eq120}
\end{equation}
Here we have dropped the Dirac delta function, which plays no role when solving Eq.~(\ref{s3:eq110}) for $E > 0$.

\begin{figure}
  \includegraphics{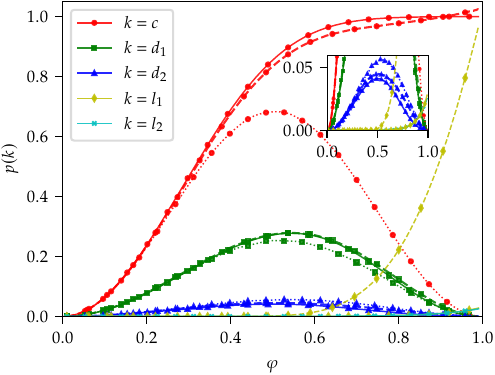}
  \caption{\label{fig-s3-10} The excitation probabilities $p_k$ as a function of $\varphi$, corresponding to a pair of Gaussian pulses with $t_0=32$ and $\varphi \in [0, 1]$. The solid curves correspond to the exact solution, while the dashed curves correspond to the effective theory. The inset shows the zoom-in for $p_k<0.06$. The probabilities from the effective theory without the high-energy correlation are plotted by the dotted curves. }
\end{figure}

In Fig.~\ref{fig-s3-10}, we compare the excitation probabilities calculated from the effective theory (dashed curves) to the exact solution (solid curves), corresponding to a pair of Gaussian pulse with $t_0=32$ and $\varphi \in [0, 1]$. Let us first concentrate on the exact solution, which are plotted by the solid curves. In this case, the excitation is dominated by the first five eh pairs, which are labeled with $c$, $d_1$, $d_2$, $l_1$ and $l_2$. The nature of these eh pairs has been discussed in our previous work \cite{Yin_2025}, which we briefly recap below. The wave function for the first three ones ($c$, $d_1$ and $d_2$) can be delocalized in the time domain: The first one $c$ is composed of the single electron and hole, which are injected by the two pulses individually. Its wave function is delocalized for $\varphi < 1.0$, but can evolve into a localized one for $\varphi=1.0$. The rest two ones $d_1$ and $d_2$ belong to the neutral cloud of eh pairs and hence are not responsible for the net charge injected by a single pulse. Their wave functions are always delocalized, but their probabilities drop to zero for $\varphi=1.0$. The last two eh pairs $l_1$ and $l_2$ are also neutral eh pairs, but with wave functions always localized in the time domain. Their probabilities $p(l_1)$ and $p(l_2)$ are strongly suppressed for pulses with noninteger flux quantum. For the Gaussian pulses with $t_0=32$, $p(l_1)$ and $p(l_1)$ almost vanish for $\varphi<0.5$, which can be better seen from the inset of Fig.~\ref{fig-s3-10}. Note that $p(l_1)$ and $p(l_2)$ are almost degenerated. This is because they are bonding-like and anti-bonding-like states built from two localized neutral eh pairs, which are injected by the two pulses individually. For $t_0 \gg 1$, the two localized eh pairs become decoupled and hence the probability difference between the bonding-like and anti-bonding-like states tends to vanish.

By comparing the dashed curves to the corresponding solid curves, one can see that the probabilities of the delocalized neutral pairs $p(d_1)$ and $p(d_2)$ can be well described by the effective theory. The effective theory can also give a good estimation for the probability $p(c)$, whose wave function undergoes the LD transition as $\varphi$ increases from $0.0$ to $1.0$. Note that $p(c)$ from the effective theory can be slightly larger than $1.0$. It is because the effective scattering matrix from Eq.~(\ref{s3:eq70}) is non-unitary. However, the effective theory fails to reproduce the main features for the localized neutral pairs. The localized excitation is dominated by only one eh pair $l_1$ from the effective theory. Its probability $p(l_1)$ is much larger than the exact result, as illustrated in Fig.~\ref{fig-s3-10} by the yellow dashed curve with diamonds.

The above results suggest that the delocalized eh pairs are dominated by the IR-divergent part of the scattering matrix, which is included exactly in the effective scattering matrix given in Eq.~(\ref{s3:eq70}). In contrast, the localized eh pairs are dominated by the high-energy correlation, which is only approximately incorporated in the effective theory by the constant term $B_0$. To justify this, we calculate the probabilities by using the effective theory from Eqs.~(\ref{s3:eq70}), (\ref{s3:eq110}) and (\ref{s3:eq120}) while setting $B_0=0$ in Eq.~(\ref{s3:eq70}). In doing so, we find the localized neutral pairs from the effective theory vanish, leaving only three delocalized pairs. The probabilities of the three delocalized pairs are plotted by dotted curves in Fig.~\ref{fig-s3-10}. By comparing to the corresponding probabilities from the exact solution, one can see that two of them represent the probabilities for the neutral pairs $d_1$ and $d_2$, as illustrated by the dotted curves with green squares and blue triangles. This indicates that the delocalized neutral pairs are indeed dominated by the IR-divergent part of the scattering matrix. One can also notice that in the absence of the high-energy correlation, the effective scattering matrix essentially corresponds to a pair of delta pulses, as we have discussed below Eq.~(\ref{s3:eq40}). It has been shown that the IR divergence in such scattering matrix is responsible for the dynamical orthogonality catastrophe, which can lead to a logarithmic divergence in the charge fluctuation \cite{levitov_1996_elect}. Hence the excitation of the delocalized eh pairs are always accompanied by the occurrence of the dynamical orthogonality catastrophe in such system.

However, the eh pair $c$ cannot be properly described without the high-energy correlation. The probability $p(c)$ evaluated from the effective theory without $B_0$ (red dotted curve) only agrees with the exact result (red solid curve) for $\varphi< 0.4$. As $\varphi$ further increases, $p(c)$ from the exact solution approaches $1.0$, while $p(c)$ from the effective theory without $B_0$ drops rapidly to zero. Indeed, as the corresponding wave function undergoes an LD transition when $\varphi$ increases from $0.0$ to $1.0$, both the high-energy correlation and low-energy divergence are needed to properly describe this eh pair.

\begin{figure}
  \includegraphics{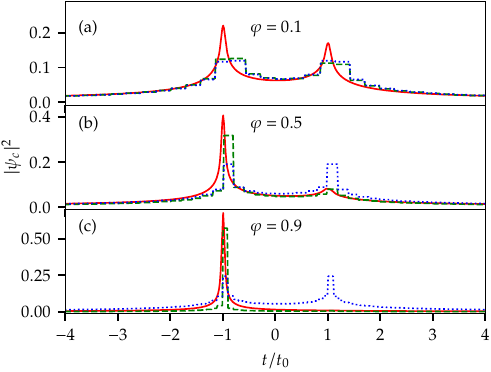}
  \caption{\label{fig-s3-20} Modulus square of wave function $|\psi_c|^2$ for $\varphi=0.1$ (a), $0.5$ (b) and $0.9$ (c), corresponding to a pair of Gaussian pulses with $t_0=32$. The red solid curves correspond to the exact solution, while the green dashed curves correspond to the effective theory. The blue dotted curves represents the wave functions evaluated from the effective theory without the high-energy correlation. }
\end{figure}

To further clarify this, we show the typical behavior of the electron wave function for the eh pair $c$ in Figs.~\ref{fig-s3-20}(a), \ref{fig-s3-20}(b) and \ref{fig-s3-20}(c), corresponding to $\varphi=0.1$, $0.5$ and $0.9$, respectively. Let us first concentrate on the wave functions from the exact result, which are represented by the red solid curves. For $\varphi=0.1$, the wave function exhibits two broad peaks around $t=\pm t_0$, indicating its delocalized nature. As $\varphi$ increases to $0.5$, the right peak is strongly suppressed, while the left peak is enhanced. As $\varphi$ reaches $1.0$, the right peak totally vanishes, leaving only a single sharp peak representing a localized wave function around $t=-t_0$. These features can be well described by the effective theory, which is represented by the green dashed curves in the figure. Note that the effective theory produces step-discretized wave functions. This is because the wave function in the effective theory can only be evaluated in the energy domain below the ultraviolet cutoff $\Lambda$. Hence the short-time information is absent for timescales smaller than $2\pi/\Lambda$. The effective theory without $B_0$ can only describe the long-time behavior of the wave function for $\varphi=0.1$. As $\varphi$ increases to $0.5$ and $1.0$, the wave functions always exhibit two broad peaks, indicating that they are always delocalized. Hence the LD transition cannot be explained, not even qualitatively, by the effective theory without the high-energy correlation.

The effective theory not only gives a qualitative estimation for the long-time behavior of the wave function $\psi_c$ but can also quantitatively describe its LD transition. To see this, we perform the single-parameter scaling analysis for the wave function from the effective theory. Following our previous work \cite{Yin_2025}, we choose the inverse participation ratio (IPR) as the scaling variable. It can be related to the wave function $\psi_c$ as
\begin{equation}
  P_c = \frac{1}{t_l} \int_{\rm{box\: origins}}  \sum_{\rm{box}(t_l)} \Big( \int_{t \in \rm{box}(t_l)} \left| \psi_c \right|^2  \Big)^2, \label{s3:eq130}
\end{equation}
where we assume the wave function is normalized in the whole time domain $\int^{+t_{\rm max}/2}_{-t_{\rm max}/2} dt \left| \psi_c \right|^2 = 1$ with $t_{\rm max} \to +\infty$. The symbol $\sum_{\rm{box}(t_l)}$ represents the summation over small boxes with a linear size $t_l$ into which one divides the whole time domain $[-t_{\rm max}/2, +t_{\rm max}/2]$. The choice of the box origins is arbitrary, which gives different values of IPR. So the IPR is further averaged over different box origins ($\frac{1}{t_l} \int_{\rm{box\: origins}} \dots$) to avoid this ambiguity.

\begin{figure}
  \includegraphics{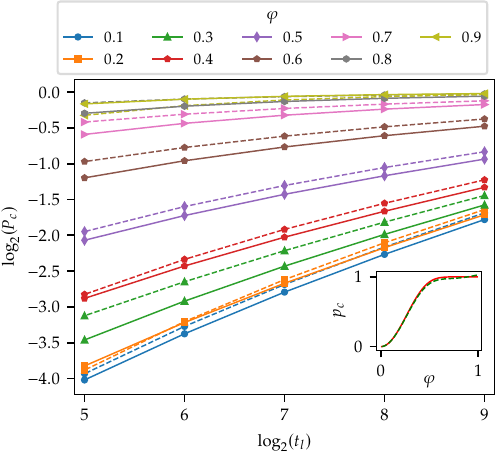}
  \caption{\label{fig-s3-30} The logarithm of the IPR $\log_2(P_c)$ as a function of logarithm of the box size $\log_2(t_l)$, corresponding to a pair of Gaussian pulses with $t_0=512$ and $t_{\rm max}=8192$. Curves with different colors and markers correspond to different value of $\varphi$. The solid curves correspond to the exact solution, while the dashed curves correspond to the effective theory. The excitation probability $p_c$ from the exact solution (red solid) and (green dashed) are shown in the inset. }
\end{figure}

To perform the scaling analysis, the IPR should be evaluated for large $t_0$ to minimize finite-size effect. We choose $t_0=512$ and $t_{\rm max}=8192$, which is proved to be suitable in our previous work. For such a large value of $t_0$, the effective theory can give a slightly better estimation of the excitation probability $p_c$. This can be seen by comparing the red solid curve (exact solution) to the green dashed curve (effective theory) in the inset of Fig.~\ref{fig-s3-30}. The typical behavior of the IPR $P_c$ is demonstrated on log-log scale in the main panel of Fig.~\ref{fig-s3-30}. Curves with different colors and markers correspond to different values of $\varphi$. Let us first concentrate on the IPR from the exact solution, which are plotted by the solid curves. One can see that the logarithm of the IPR $\log_2(P_c)$ increases almost linearly as a function of $\log_2(t_l)$ when $\varphi$ is small, indicating the corresponding wave function $\psi_c$ is delocalized in the time domain. As $\varphi$ approaches $1.0$, $\log_2(P_c)$ tends to be independent on $t_l$, corresponding to a localized wave function. The different box size dependence of the IPR thus display a typical signature of the LD transition. The IPR from the effective theory are plotted by dashed curves in the main panel of Fig.~\ref{fig-s3-30}. The dashed curves are almost parallel to the corresponding solid curves. This suggests that the effective theory can at least qualitatively reproduce the main feature of the LD transition.
\begin{figure}
  \includegraphics{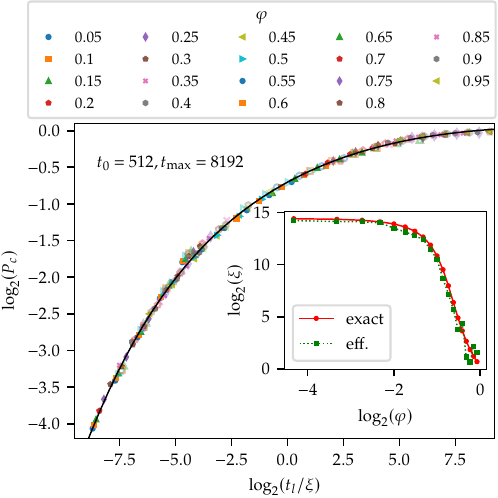}
  \caption{\label{fig-s3-40} The IPR $\log_2(P_c)$ as a function of $\log_2(t_l/\xi)$, corresponding to a pair of Gaussian pulses with $t_0=512$ and $t_{\rm max}=8192$. The filled markers correspond to the exact solution, while the unfilled markers correspond to the effective theory. The black solid curve represents the scaling function. The correlation length $\log_2(\xi)$ as a function of $\log_2(\varphi)$ is demonstrated in the inset. The red solid curve corresponds to the exact solution, while the green dotted curve corresponds to the effect theory. }
\end{figure}

Given the IPR, we then perform the single-parameter scaling analysis to quantitatively characterize the LD transition. We assume the IPR $P_c$ for a given $t_0$ can be described by a scaling function $f(x)$ as
\begin{equation}
  P_c(t_l, \varphi) = f[t_l/\xi(\varphi)],
  \label{s3:eq140}
\end{equation}
with $\xi(\varphi)$ representing the correlation length. Both the scaling function and the correlation length can be obtained by using the data collapse method. To do this, we rescale the IPR $P_c$ for various $t_l$ and $\varphi$ so that all the data points can be collapsed into a single curve, corresponding to the scaling function $f(x)$. In the data collapse analysis for the exact solution, we use the IPR evaluated for box size $t_l \in [32, 512]$ \footnote{This region is smaller than the region $\varphi \in [16, 512]$, which we have used in Ref.~\cite{Yin_2025}. This has little effect on the scaling function and correlation length, but can give a slightly better quality of the data collapse.}. For the effective theory, we use the IPR evaluated for the box size $t_l \in [64, 512]$, as we notice from Fig.~\ref{fig-s3-30} that the IPR from the effective theory drops slightly faster than the IPR from the exact solution when $t_l < 64$. From the main panel of Fig.~\ref{fig-s3-30}, one can see that this effect is more pronounced for $\varphi=0.2$, $0.4$ and $0.9$. We believe that this is because the effective theory lacks short-time information of the wave function due to the finite cutoff, which has been demonstrated in Fig.~\ref{fig-s3-20}. The collapse of the IPR is demonstrated on log-log scale in the main panel of Fig.~\ref{fig-s3-40}. The filled markers correspond to the exact solution, while the unfilled markers correspond to the effective theory. One can see that both of them can be well collapsed into the same curve, which is shown by the black solid curve in the main panel. So the effective theory can give the same scaling function $f(x)$ as the one from the exact solution.

The correlation lengths $\xi$ as a function of $\varphi$ are plotted on log-log scale in the inset of Fig.~\ref{fig-s3-40}. The red solid curve corresponds to the exact solution. It diverges as $\varphi$ decreases from $1.0$ to $0.0$, which is a key signature of the phase transition. The divergence is rounded when $\xi$ becomes comparable to the system size, which can be attributed to the finite-size effect. These features can also be captured by the effective theory, whose correlation length is plotted by the green dotted curve. Note that the correlation length from the effective theory can exhibit a small oscillation. The oscillation becomes pronounced when $\varphi$ is close to $1.0$. In this region, the scattering matrix becomes dominated by the high-energy correlation, which is only approximated described by the constant term $B_0$ in Eq.~(\ref{s3:eq70}). So we expect that the oscillation is induced by the inaccuracy of the effective scattering matrix in this region.

\begin{table}
  \caption{Voltage pulses}
  \begin{ruledtabular}
    \begin{tabular}{ll}
      Name & Expression \\
      Gaussian & $V_p(t) = \frac{\varphi}{\sqrt{\pi\ln(2)}} \exp[ -\ln(2)t^2 ]$\\
      Hyperbolic secant & $V_p(t) = \frac{\varphi}{\pi} \ln(2+\sqrt{3}) \sech[\ln(2+\sqrt{3})t]$\\
      Lorentzian-squared & $V_p(t) = \frac{\varphi}{\pi^2} \frac{1}{\left( t^2+1 \right)^2}$\\      
    \end{tabular}
  \end{ruledtabular}
  \label{s3:tab10}
\end{table}

\begin{figure}
  \includegraphics{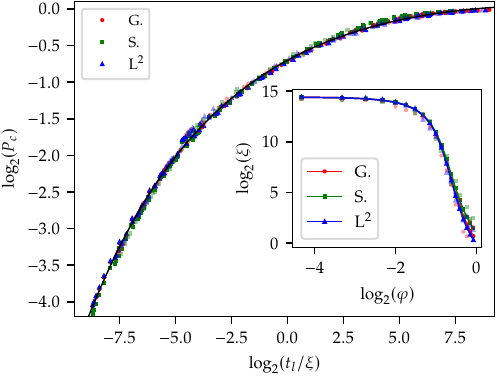}
  \caption{\label{fig-s3-50} The IPR $\log_2(P_c)$ as a function of $\log_2(t_l/\xi)$. The red dots, green squares and blue triangles correspond to the Gaussian (G), hyperbolic secant (S), Lorentzian-squared (L2) pulses, respectively. The filled markers represent the IPR from the exact solution, while the unfilled markers represent the IPR from the effective theory. }
\end{figure}

From the above discussion, one can see that the effective theory can reproduce both the scaling function and correlation length of the LD transition for the Gaussian pulse. Now let us check if it also works for other short-tailed pulses. To show this, we compare the LD transition for three different voltage pulses as shown in Table~\ref{s3:tab10}. The collapse of the IPR as a function of $t_l/\xi$ are demonstrated on log-log scale in the main panel of Fig.~\ref{fig-s3-50}, corresponding to $t_0=512$ and $t_{\rm max}=8192$. The red dots, green squares and blue triangles correspond to the Gaussian (G), Hyperbolic secant (S), and Lorentzian-squared (L2) pulses, respectively. The IPR from the exact solution are plotted by the filled markers, while the IPR from the effective theory are plotted by unfilled markers. One can see that all the data points are collapsed into the same scaling function, which is illustrated by the black solid curve in the main panel. In the meantime, the corresponding correlation lengths also diverge in a similar manner, as illustrated in the inset. These results show that the LD transition for these pulses belong to the same universality class, which is described by the effective scattering matrix given in Eq.~(\ref{s3:eq70}). To make the effective theory applicable, it is not necessary to require the Taylor expansion in Eq.~(\ref{s3:eq50}) converges order by order. It works as long as the high-energy correlation can be approximated by the constant term $B_0$ given in Eq.~(\ref{s3:eq105}). This occurs for the case of the Lorentzian-squared ($L^2$) pulse, as illustrated by the blue triangles in Fig.~\ref{fig-s3-50}.

\section{Low-energy scattering theory for long-tailed pulse \label{sec4}}

The effective theory developed in the previous section only works for pulses with sufficient short tails. In this case, the IR divergence of the scattering matrix is dominated solely by the $1/E$ term in Eq.~(\ref{s3:eq70}), which is also responsible for the excitation of delocalized eh pairs. For pulses with long tails, the scattering matrix can exhibit additional IR divergence. This can lead to the LD transition belonging to different types of university class. 

\subsection{Lorentzian pulses \label{sec4.1}}

As a typical example, let us consider a pair of Lorentzian pulses. In this case, the pulses decays as $\sim t^{-2}$ at long times. The integral in Eq.~(\ref{s3:eq105}) does not converge, indicating the presence of additional IR divergences. To further explore this, we rewrite Eqs.~(\ref{s3:eq10}),~(\ref{s3:eq20}) and~(\ref{s3:eq30}) into
\begin{equation}
  S(E) = e^{-i E t_0} S_p(E) + e^{i E t_0} S_p(-E).
  \label{s4:eq10}
\end{equation}
Now $S_p(E)$ represents the scattering matrix of a single Lorentzian pulse, which can be written as
\begin{equation}
  S_p(E) = \int^{t_0}_{-\infty} dt e^{-i E t} e^{ - 2\pi i \phi_p(t)},
  \label{s4:eq20}
\end{equation}
where the scattering phase $\phi_p(t)$ can be written as $\phi_p(t) = 2 \varphi \arctan(t)$.

In the limit $t_0 \to +\infty$ and $E \to E + i \eta$, the scattering matrix $S_p(E)$ from Eq.~(\ref{s4:eq20}) can be evaluated analytically by using the contour integral technique, which gives
\begin{equation}
  S_p(E) = \frac{2 \sin(\pi \varphi)}{E} e^{-E} \Gamma(1-\varphi) U(-\varphi, 0, 2E),
  \label{s4:eq30}
\end{equation}
with $U(a, b, c)$ representing the confluent hypergeometric function of the second kind and $\Gamma(x)$ representing the Gamma function \cite{Abramowitz_1965}. By using {\it Mathematica}, it can be expanded in a power series of $E$ for $E>0$ as
\begin{equation}
  S_p(E) = \frac{2 \sin(\pi \varphi)}{E} + B_1 \ln(E) + B^{+}_0 + O(E),
  \label{s4:eq32}
\end{equation}
with
\begin{eqnarray}
C_0 & = & 2 \sin(\pi \varphi), \nonumber\\
  B^+_0 & = & -2 \sin(\pi\varphi) \Big( 1 - 2\varphi + 4\varphi\gamma \nonumber\\
      &&{}+ 2\varphi \ln(2) + 2\varphi \digamma[1-\varphi] \Big), \\
  B_1 & = &-2 \varphi \sin(\pi\varphi),
            \label{s4:eq35}
\end{eqnarray}
where $\digamma(x)$ represents the digamma function and $\gamma$ represents the Euler constant \cite{Abramowitz_1965} .

In analogy to the previous section, we choose the effective scattering matrix by truncating the series expansion up to the zeroth order of $E$. This gives
\begin{equation}
  e^{-i \pi \varphi} S^{\rm eff}_p(E) =  C_0 \frac{e^{i E t_0} - 1}{E} + B_1 \ln(E) + B^{+}_0.
  \label{s4:eq40}
\end{equation}
In doing so, the electron number $N(E)$ from the effective theory can be expressed as in the limit  $t_0 \to +\infty$ and $E \to E + i \eta$:
\begin{equation}
  N(E) = -\frac{C^2_0}{2\pi} \ln(E) + \frac{C_0 B_1}{2\pi} E \ln^2(E) + \frac{C_0 (B^{+}_0-B_1)}{\pi} E \ln(E) + N_0.
  \label{s4:eq50}
\end{equation}
It can properly reproduce the non-analytical behavior of the electron number $N(E)$ from the exact solution, despite the choice of the cutoff $\Lambda$. Note that  the electron number $N(E)$ of the Lorentzian pulse contains an additional non-analytic term $(C_0 B_1/2\pi) E \ln^2(E)$, which is absent in the case of short-tailed pulses [Eq.~(\ref{s3:eq100})]. This can give rise to a logarithm divergence in the electron distribution function $f(E)=\partial_E N(E)$, which has been addressed in the previous study \cite{moskalets_2016_fract}.

While the non-analytical terms in Eq.~(\ref{s4:eq50}) are independent on $\Lambda$, the constant $N_0$ is sensitive to $\Lambda$, which has the form
\begin{eqnarray}
  N_0 & = & \frac{C^2_0}{2\pi} [\ln(\Lambda) -\ln(2)] + \frac{C_0B^+_0}{\pi}2\ln(2)\Lambda \nonumber\\
      &&{} - \frac{C_0B_1}{2\pi}\Lambda[\ln^2(\Lambda) - \Lambda \ln^2(\Lambda) + 2\ln^2(2)- 4\ln(2)] \nonumber\\
      &&{} + \Big[ \frac{(B^+_0)^2}{2\pi} + \frac{B^2_1}{\pi} - \frac{B^{+}_0B_1}{\pi} \Big] \Lambda^2 \nonumber\\
      &&{} + \frac{(B^+_0)^2}{2\pi} \Lambda^2 [ 2\ln^2(2) -2\ln(2) + \frac{3}{2} + \ln^2(\Lambda) - \ln(\Lambda)]  \nonumber\\
      &&{} + \Big( \frac{B^2_1}{\pi} - \frac{B^+_0B_1}{\pi} \Big) [ \frac{1}{2} - 2\ln(2) - \ln(\Lambda) ] \Lambda^2.
  \label{s4:eq60}
\end{eqnarray}
By requiring the constant term $N_0$ is equal to its counterpart from the exact solution, one obtains the value of the cutoff $\Lambda$.

To evaluate the wave function and excitation probability, we need to substitute the effective scattering matrix of a single Lorentzian pulse $S_p(E)$ [Eq.~(\ref{s4:eq40})] into Eqs.~(\ref{s3:eq110}) and (\ref{s3:eq120}). However, this requires $S_p(E)$ for $E<0$, which is invalid from Eq.~(\ref{s4:eq40}). To fix this, we evaluate the effective scattering matrix for $E<0$ separately, which gives
\begin{equation}
  e^{-i \pi \varphi} S^{\rm eff}_p(E) =  C_0 \frac{e^{i E t_0} - 1}{E} + B_1 \ln(-E) + B^{-}_0,
  \label{s4:eq80}
\end{equation}
for $E<0$. The coefficient $B^-_0$ can be written as
\begin{eqnarray}
  B^-_0 & = & -2 \sin(\pi\varphi/2) \Big( -1 - 2\varphi + 4\varphi\gamma \nonumber\\
      &&{}+ 2\varphi \ln(2) + 2\varphi \digamma[1+\varphi] \Big).
            \label{s4:eq90}
\end{eqnarray}
\begin{figure}
  \includegraphics{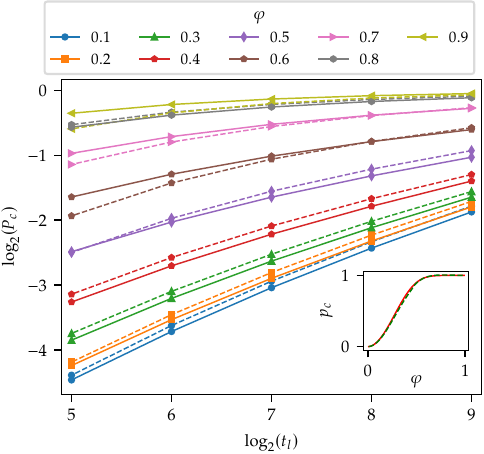}
  \caption{\label{fig-s4-10} The same as Fig.~\ref{fig-s3-30}, but for the Lorentzian pulse. }
\end{figure}

By properly account for the additional logarithmic divergence, the effective theory can give a good estimation for both the excitation probability and the IPR for the Lorentzian pulse. This is demonstrated in Fig.~\ref{fig-s4-10}, corresponding to $t_0=512$ and $t_{\rm max}=8192$. The excitation probability $p_c$ is plotted in the inset. The red solid and green dashed curves correspond to the exact solution and effective theory, respectively. They show a good agreement with each other. The corresponding IPR $P_c$ is demonstrated on log-log scale in the main panel. Curves with different colors and markers correspond to different values of $\varphi$. The solid curves represent the IPR from the exact solution, which shows a clear signature of the LD transition. The dashed curves represent the IPR from the effective theory. Similar to the case of short-tailed pulses, they exhibit similar behaviors as the ones from the exact solution when the box size $t_l$ is not too small ($t_l > 64$). For $t_l<64$, the IPR from the effective theory drops more rapidly than the one from the exact solution, which we attributed to the loss of accuracy at short times.

\begin{figure}
  \includegraphics{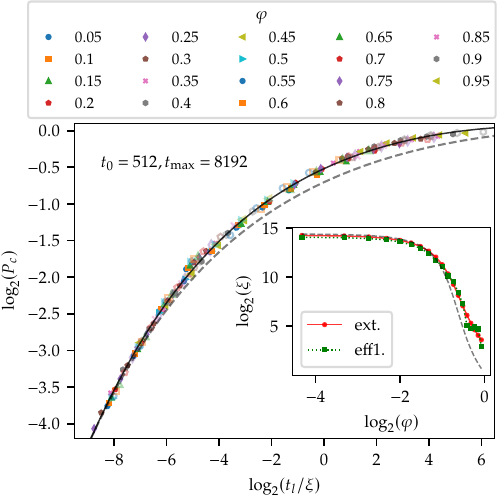}
  \caption{\label{fig-s4-20} The same as Fig.~\ref{fig-s3-40}, but for the Lorentzian pulse. The black dashed curves in the main panel and inset represent the scaling function and correlation length for the Gaussian pulse, respectively. }
\end{figure}

Now we perform the single-parameter scaling analysis by using the data collapse method following Eq.~(\ref{s3:eq140}). In the data collapse analysis for the exact solution, we use the IPR evaluated for box size $t_l \in [32, 512]$. For the effective theory, we use the IPR evaluated for the box size $t_l \in [64, 512]$. The collapse of the IPR is demonstrated on log-log scale in the main panel of Fig.~\ref{fig-s4-20}. The filled markers correspond to the exact solution, while the unfilled markers correspond to the effective theory. By rescaling the IPR, we find that they can be collapsed into the same black solid curve, as shown in the main panel. The corresponding correlation lengths are also plotted in the inset. The red solid and green dashed curves correspond to the exact solution and effective theory, respectively. They also show a good agreement. These results show that the effective scattering matrix from Eqs.~(\ref{s4:eq40}) and~(\ref{s4:eq80}) can be used to quantitively characterize the LD transition for the Lorentzian pulse. 

In the main panel of Fig.~\ref{fig-s4-20}, we plot the scaling function for the Gaussian pulse by the black dashed curve. The corresponding correlation length is also plotted by the black dashed curve in the inset. Comparing to the case of the Lorentzian pulse, both the scaling function and correlation length show clear different behaviors, indicating that they belong to different university classes. The difference can be attributed to the additional logarithmic divergence of the scattering matrix given in Eqs.~(\ref{s4:eq40}) and~(\ref{s4:eq80}). It is present for the Lorentzian pulse, but absent for pulses with short tails [Eq.~(\ref{s3:eq70})].

\subsection{Fractional-powered Lorentzian pulse\label{sec4.2}}

What happens when the voltage pulse decays slower than the Lorentzian? To explore this, we consider a family of pulses, whose temporal profile can be written as
\begin{equation}
V_p(t) =  \frac{\varphi\Gamma(\alpha)}{\sqrt{\pi}\Gamma(\alpha-1/2)} \frac{1}{\left( t^2+1 \right)^{\alpha}}.
\label{s4:eq100}
\end{equation}
The parameter $\alpha$  satisfies $\alpha > 0.5$ so that the voltage pulse carries a finite flux $\varphi = \int dt V_p(t)$. 

The above expression corresponds to the Lorentzian pulse with an arbitrary exponent, which decays as $\sim t^{-2\alpha}$ at long times. For $\alpha > 1.0$, it is sufficiently sharp so that it can be well approximated by the effective scattering matrix from Eq.~(\ref{s3:eq70}). The corresponding scattering matrix exhibits only $1/E$ divergence in the IR limit. A specific case occurs for $\alpha=2.0$, corresponding to the Lorentzian-squared pulse discussed in the end of Sec.~\ref{sec3}. For $\alpha=1.0$, it reduces to the Lorentzian, corresponding to the scattering matrix with an additional logarithmic divergence. For $\alpha < 1.0$, it is a fractional-powered Lorentzian pulse. The scattering matrix can exhibit a complicated divergence, which we do not know how to handle in a general way. In this paper, we concentrate on a more restricted parameter space $\alpha \in (0.75, 1.0)$, where the IR divergence can be obtained analytically. To see this, we decompose the scattering matrix of a single pulse as
\begin{eqnarray}
    S_p(E) &  = &  (e^{-2\pi i \varphi} - 1) \frac{e^{i E t_0} - 1}{i E} \nonumber\\
    &&\hspace{-1.25cm}{}+ 4\pi e^{-i\pi\varphi} {\rm Re}\Big[ e^{i\pi\varphi} \int^{0}_{-\infty} dt V_p(t) e^{-i\phi_p(t)}  \frac{e^{i E t} - 1}{E} \Big],
         \label{s4:eq110}
\end{eqnarray}
where $\phi_p(t)$ represents the forward scattering phase due to a single pulse. In the above expression, the first term in the right-hand side gives the $1/E$ divergence, which lead to the orthogonality catastrophe. The second term can give an additional IR divergence, which is slower than $1/E$. The divergence can be obtained by using the series expansion for $0.75 < \alpha < 1.0$, which gives
\begin{equation}
    \int^{0}_{-\infty} dt V_p(t) e^{-i\phi_p(t)}  \frac{e^{i E t} - 1}{E} = \varphi D^{\pm}_\alpha E^{2\alpha-1} + D_0 + O(E).
    \label{s4:eq120}
\end{equation}
The coefficient $D^{\pm}_\alpha$ takes different values for $E>0$ and $E<0$, it can be written as
\begin{equation}
  D^{\pm}_\alpha = \frac{\Gamma(1/2+\alpha)\Gamma(1/2-\alpha) \mp i \pi \sec(\pi\alpha-\pi/2)}{2^{2\alpha-1}\Gamma(\alpha-1/2)\Gamma(1/2+\alpha)},
    \label{s4:eq121}
\end{equation}
with $D^+_\alpha$ and $D^-_\alpha$ corresponding to $E>0$ and $E<0$, respective. The constant term $D_0$ is independent on the sign of $E$, which can be evaluated numerically as
\begin{equation}
  D_0 = \lim_{E \to 0} \Big[\int^{0}_{-\infty} dt V_p(t) (e^{i E t} - 1) -  \varphi D^{\pm}_\alpha E^{2\alpha-1}\Big].
    \label{s4:eq122}
\end{equation}
The above expansion suggests that the scattering matrix can exhibit an additional fractional power-law divergence. The low-energy effective theory can be constructed following the similar procedure introduced in the previous section. Up to the zeroth-order of the energy $E$, we find that the effective scattering matrix for the fractional-powered Lorentzian pulse can be written as
\begin{equation}
e^{-i \pi \varphi} S^{\rm eff}_p(E) = \frac{C_0}{E} + B^{\pm}_1 \frac{|E|^{2\alpha-1}}{E} + B_0,
    \label{s4:eq130}
\end{equation}
with $C_0$, $B_0$ and $B^{\pm}_1$ being real coupling parameters. By using the perturbative matching, we obtain
\begin{eqnarray}
  C_0 & = & 2 \sin(\pi \varphi), \\
  B^{\pm}_1 & = &2 \varphi {\rm Re}(e^{i\pi\varphi}D^{\pm}_\alpha) , \\
  B_0 & = &2  {\rm Re}(D_0).
            \label{s4:eq140}
\end{eqnarray}
They are independent on the ultraviolet cutoff $\Lambda$. The cutoff $\Lambda$ is obtained by matching the non-divergent part of the electron number $N(E)$ in the limit $E\to 0$.

\begin{figure}
  \includegraphics{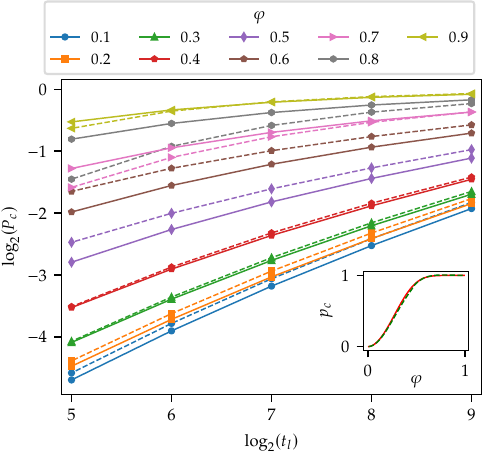}
  \caption{\label{fig-s4-30} The same as Fig.~\ref{fig-s3-30}, but for the fractional-powered Lorentzian pulse with $\alpha=0.85$. }
\end{figure}

In Fig.~\ref{fig-s4-30}, we plot the IPR $P_c$ and the excitation probability $p_c$ for the fractional-powered Lorentzian pulse, corresponding to $\alpha=0.85$, $t_0=512$ and $t_{\rm max}=8192$. The excitation probability $p_c$ is shown in the inset. The red solid and green dashed curves correspond to the exact solution and effective theory, respectively. One can see that the effective theory can well reproduce the $\varphi$ dependence of the probability. The IPR $P_c$ are plotted in the main panel. Curves with different colors and markers correspond to different values of $\varphi$. The solid curves represent the IPR from the exact solution, while the dashed curves represent the IPR from the effective theory. One finds that the effective theory can also give a good estimation for the IPR when $t_l$ is not too small ($t_l > 64$).

\begin{figure}
  \includegraphics{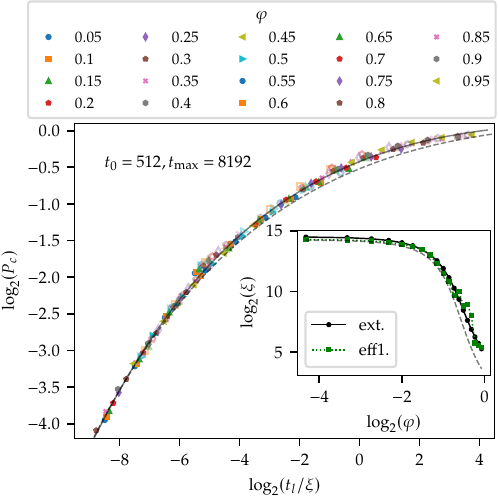}
  \caption{\label{fig-s4-40} The same as Fig.~\ref{fig-s3-40}, but for the fractional-powered Lorentzian pulse with $\alpha=0.85$. The black dashed curves in the main panel and inset represent the scaling function and correlation length for the Lorentzian pulse, respectively. }
\end{figure}

Given the IPR, we then perform the single-parameter scaling analysis by using the data collapse method following Eq.~(\ref{s3:eq140}). In the data collapse analysis for the exact solution, we use the IPR evaluated for box size $t_l \in [32, 512]$. For the effective theory, we use the IPR evaluated for the box size $t_l \in [64, 512]$. The collapse of the IPR is demonstrated on log-log scale in the main panel of Fig.~\ref{fig-s4-40}, while the corresponding correlation length is shown in the inset. In the main panel, the filled markers represent the IPR from the exact solution, while the unfilled markers represent the IPR from the effective theory. By properly rescaling the IPR, the data points from both the exact solution and effective theory can be collapsed into the same black solid curve, indicating that the effective theory can reproduce the scaling function of the LD transition. In the meantime, the effective theory can also give a good estimation for the correlation length. This can be seen by comparing the red solid curve to the green dotted one in the inset. Hence we can conclude that the LD transition in this case is closely related to the additional fractional power-law divergence in the IR limit, which is emphasized by the second term in the effective scattering matrix Eq.~(\ref{s4:eq130}). Due to the contribution of this divergence, both the scaling function and the correlation length exhibit quantitatively different behaviors from the ones of the Lorentzian pulse, as illustrated by the black dashed curves in the main panel and inset, respectively.

\begin{figure}
  \includegraphics{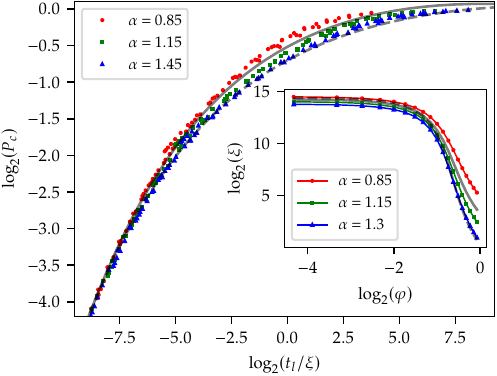}
  \caption{\label{fig-s4-50} The IPR $\log_2(P_c)$ as a function of $\log_2(t_l/\xi)$ for $t_0=512$ and $t_{\rm max}=8192$. The red dots, green squares and blue triangles correspond to the fractional-powered Lorentzian with $\alpha=0.85$, $\alpha=1.15$ and $\alpha=1.45$, respectively. The black solid (dashed) curves in the main panel and inset represent the scaling function and correlation length for the Lorentzian (Gaussian) pulse, respectively }
\end{figure}

The above results suggest that the LD transition can be manipulated by changing the profile of the voltage pulse. In particular, as the exponent $\alpha$ increases from below to above $1.0$, the voltage pulse evolves smoothly from a long-tailed one to a short-tailed one, which correspond to LD transitions belonging to different university classes. This is demonstrated in Fig.~\ref{fig-s4-50}. The collapse of the IPR are shown in the main panel. They are collapsed into three different curves representing the corresponding scaling functions. The scaling function for $\alpha=0.85$ (red dots) is apparently different from the rest two ones, which lies well above the scaling function for the Lorentzian pulse, as illustrated by the black solid curve. In contrast, the scaling function for $\alpha=1.15$ (green squares) lies below the scaling function for the Lorentzian pulse, but is still slightly higher than the scaling function for the Gaussian pulse, as illustrated by the black dashed curve. The scaling function for $\alpha=1.45$ nearly coincides with the scaling function for the Gaussian pulse, as can be seen by comparing the blue triangles to the black dashed curve. A similar behavior can also be seen for the correlation length, which is shown in the inset. These results show that critical behavior of LD transition can be fine tuned by using $\alpha$ as a tuning parameter.

\section{Conclusion \label{sec5}}

In this paper, we present a low-energy effective theory, which can be used to describe the LD transition for the wave function of electrons and holes injected individually by a single voltage pulse. The LD transition is induced by the competition between two different parts in the scattering matrix. The first part describes the IR divergence of the scattering matrix, while the second part represents the high-energy correlation. At the low-energy space below a given cutoff, we find that the high-energy correlation can always be well approximated by a constant, but the IR divergence has to be described exactly. For short tailed pulses which decay slower than Lorentzian, we show that the scattering matrix exhibits only an inverse linear divergence in the IR limit, which is only decided by the Faraday flux of the voltage pulse. The divergence is also responsible for the dynamical orthogonality catastrophe. Hence the LD transition for all short-tailed pulses can be described by the same effective scattering matrix, leading to the LD transition belonging to the same university class. In contrast, the scattering matrix can exhibit additional divergences for long-tailed pulses, such as a logarithmic divergence for the Lorentzian pulse and a power-law divergence for the fractional-powered Lorentzian pulses. This can lead to the LD transition belonging to different university classes. This provides a mechanism to manipulate the LD transition in such system.

\appendix*
\section{Details of the perturbative matching}

In this appendix, we present some technical details of the perturbative matching. Let us first concentrate on the short-tailed pulse, which is discussed in Sec.~\ref{sec3}. The electron distribution function $f(E)$ from the exact scattering matrix can be obtained by substituting Eq.~(\ref{s3:eq40}) into Eq.~(\ref{s3:eq90}) as $S^{\rm eff}(E)$. As we focus on the limit $t_0 \to +\infty$, we introduce a small positive imaginary part to the energy $E \to E + i \eta$. By first letting $t_0 \to +\infty$ and then letting $\eta \to 0$, the $t_0$ dependence in $f(E)$ can be eliminated, which gives
\begin{equation}
    f(E) =  \int^{\Lambda+E}_{E} \frac{d\epsilon}{2\pi} | \frac{C_0}{E} + F(E)|^2.
    \label{a1:eq10}
\end{equation}
The coefficient $C_0$ and the function $F(E)$ are all real for the symmetric pulse, which can be written as
\begin{eqnarray}
    C_0 & = & 2 \sin(\pi\varphi), \\
    F(E) & = & e^{i \pi \varphi}\int^{+\infty}_{-\infty} dt 2 \pi V_p(t) e^{-i\phi_p(t)} \frac{e^{i E t} - 1}{E}.
    \label{a1:eq20}
\end{eqnarray}
By carrying out the integral in Eq.~(\ref{a1:eq10}), one obtains
\begin{equation}
    f(E) =  \frac{C^2_0}{2\pi} \frac{1}{E} - \frac{C_0}{\pi}F(E) \ln(E) + f_N(E),
    \label{a1:eq30}
\end{equation}
with
\begin{eqnarray}
f_N(E) & = &  \frac{C_0}{\pi}F(E+\Lambda) \ln(E+\Lambda) - \frac{C_0}{\pi} \int^{\Lambda+E}_E d\epsilon \ln(\epsilon) F' \nonumber\\
&&{} + \int^{\Lambda+E}_E \frac{d\epsilon}{2\pi} F^2(\epsilon) - \frac{C^2_0}{2\pi}  \frac{1}{E+\Lambda}.
    \label{a1:eq40}
\end{eqnarray}
where $F'(E)$ represents the derivative of $F(E)$. For the short-tailed pulse, the function $F(E)$ is analytical, which can be expanded into a Taylor series. As a consequence, the last term $f_N(E)$ in the right-hand side of Eq.~(\ref{a1:eq30}) is analytical, while the first two terms are non-analytical. By substituting Eq.~(\ref{a1:eq30}) into Eq.~\ref{s3:eq80}, one obtains the electron number $N(E)$:
\begin{eqnarray}
N(E) & = & -\frac{C^2_0}{2\pi} \ln(E) + \frac{C_0}{\pi} \int^E_0 d\epsilon F(\epsilon) \ln(\epsilon) \nonumber\\
&&{}+ [ \frac{C_0}{\pi} \int^0_{\Lambda} d\epsilon F(\epsilon) \ln(\epsilon) + \frac{C^2_0}{2\pi}\ln(\Lambda) + \int^{\Lambda}_E d\epsilon f_N(\epsilon) ].
    \label{a1:eq50}
\end{eqnarray}
where the first two terms in the right-hand side are non-analytical, while the last term is analytical. Note that only the last term is dependent on the cutoff $\Lambda$. The above expression works for both the exact solution and effective theory.  For the exact solution, we evaluate $F(E)$ exactly from Eq.~(\ref{a1:eq20}) and choose $\Lambda \to +\infty$. For the effective theory, we choose $\Lambda$ as an undetermined constant and set $F(E)=B_0$, corresponding to the effective scattering matrix given in Eq.~(\ref{s3:eq70}). The perturbative matching can be achieved by comparing the value and $n$-th order derivates of $N(E)$ in the limit $E \to 0$: 1) the non-analytic terms decide the parameter $B_0$; 2) the analytic term, \eg:
\begin{equation}
    N_{\rm nDOC} = \lim_{E \to 0} [N(E) + \frac{C^2_0}{2\pi} \ln(E)],
    \label{a1:eq60}
\end{equation}
gives the cutoff $\Lambda$.

The perturbative matching for the long-tailed pulses can be done in a similar way: One first splits the scattering matrix into the analytical and non-analytical parts, then treats the non-analytical part exactly, while approximates the analytical part by the constant $B_0$. In certain cases, such as the fractional-powered Lorentzian pulse discussed in Sec.~\ref{sec4.2}, the analytical expression for $N_{\rm nDOC}$ can be quite involved and one can evaluate it numerically from Eq.~(\ref{a1:eq60}).

\bibliography{refs}{}

\begin{thebibliography}{43}%
\makeatletter
\providecommand \@ifxundefined [1]{%
 \@ifx{#1\undefined}
}%
\providecommand \@ifnum [1]{%
 \ifnum #1\expandafter \@firstoftwo
 \else \expandafter \@secondoftwo
 \fi
}%
\providecommand \@ifx [1]{%
 \ifx #1\expandafter \@firstoftwo
 \else \expandafter \@secondoftwo
 \fi
}%
\providecommand \natexlab [1]{#1}%
\providecommand \enquote  [1]{``#1''}%
\providecommand \bibnamefont  [1]{#1}%
\providecommand \bibfnamefont [1]{#1}%
\providecommand \citenamefont [1]{#1}%
\providecommand \href@noop [0]{\@secondoftwo}%
\providecommand \href [0]{\begingroup \@sanitize@url \@href}%
\providecommand \@href[1]{\@@startlink{#1}\@@href}%
\providecommand \@@href[1]{\endgroup#1\@@endlink}%
\providecommand \@sanitize@url [0]{\catcode `\\12\catcode `\$12\catcode `\&12\catcode `\#12\catcode `\^12\catcode `\_12\catcode `\%12\relax}%
\providecommand \@@startlink[1]{}%
\providecommand \@@endlink[0]{}%
\providecommand \url  [0]{\begingroup\@sanitize@url \@url }%
\providecommand \@url [1]{\endgroup\@href {#1}{\urlprefix }}%
\providecommand \urlprefix  [0]{URL }%
\providecommand \Eprint [0]{\href }%
\providecommand \doibase [0]{https://doi.org/}%
\providecommand \selectlanguage [0]{\@gobble}%
\providecommand \bibinfo  [0]{\@secondoftwo}%
\providecommand \bibfield  [0]{\@secondoftwo}%
\providecommand \translation [1]{[#1]}%
\providecommand \BibitemOpen [0]{}%
\providecommand \bibitemStop [0]{}%
\providecommand \bibitemNoStop [0]{.\EOS\space}%
\providecommand \EOS [0]{\spacefactor3000\relax}%
\providecommand \BibitemShut  [1]{\csname bibitem#1\endcsname}%
\let\auto@bib@innerbib\@empty
\bibitem [{\citenamefont {Anderson}(1967)}]{Anderson1967}%
  \BibitemOpen
  \bibfield  {author} {\bibinfo {author} {\bibfnamefont {P.~W.}\ \bibnamefont {Anderson}},\ }\bibfield  {title} {\bibinfo {title} {Infrared catastrophe in fermi gases with local scattering potentials},\ }\href {https://doi.org/10.1103/physrevlett.18.1049} {\bibfield  {journal} {\bibinfo  {journal} {Physical Review Letters}\ }\textbf {\bibinfo {volume} {18}},\ \bibinfo {pages} {1049} (\bibinfo {year} {1967})}\BibitemShut {NoStop}%
\bibitem [{\citenamefont {Nozi\`eres}\ and\ \citenamefont {De~Dominicis}(1969)}]{NOZI_RES_1969}%
  \BibitemOpen
  \bibfield  {author} {\bibinfo {author} {\bibfnamefont {P.}~\bibnamefont {Nozi\`eres}}\ and\ \bibinfo {author} {\bibfnamefont {C.~T.}\ \bibnamefont {De~Dominicis}},\ }\bibfield  {title} {\bibinfo {title} {Singularities in the x-ray absorption and emission of metals. iii. one-body theory exact solution},\ }\href {https://doi.org/10.1103/physrev.178.1097} {\bibfield  {journal} {\bibinfo  {journal} {Physical Review}\ }\textbf {\bibinfo {volume} {178}},\ \bibinfo {pages} {1097} (\bibinfo {year} {1969})}\BibitemShut {NoStop}%
\bibitem [{\citenamefont {Schotte}\ and\ \citenamefont {Schotte}(1969)}]{Schotte_1969}%
  \BibitemOpen
  \bibfield  {author} {\bibinfo {author} {\bibfnamefont {K.~D.}\ \bibnamefont {Schotte}}\ and\ \bibinfo {author} {\bibfnamefont {U.}~\bibnamefont {Schotte}},\ }\bibfield  {title} {\bibinfo {title} {Tomonaga’s model and the threshold singularity of x-ray spectra of metals},\ }\href {https://doi.org/10.1103/physrev.182.479} {\bibfield  {journal} {\bibinfo  {journal} {Physical Review}\ }\textbf {\bibinfo {volume} {182}},\ \bibinfo {pages} {479} (\bibinfo {year} {1969})}\BibitemShut {NoStop}%
\bibitem [{\citenamefont {Ohtaka}\ and\ \citenamefont {Tanabe}(1990)}]{Ohtaka_1990}%
  \BibitemOpen
  \bibfield  {author} {\bibinfo {author} {\bibfnamefont {K.}~\bibnamefont {Ohtaka}}\ and\ \bibinfo {author} {\bibfnamefont {Y.}~\bibnamefont {Tanabe}},\ }\bibfield  {title} {\bibinfo {title} {Theory of the soft-x-ray edge problem in simple metals: historical survey and recent developments},\ }\href {https://doi.org/10.1103/revmodphys.62.929} {\bibfield  {journal} {\bibinfo  {journal} {Reviews of Modern Physics}\ }\textbf {\bibinfo {volume} {62}},\ \bibinfo {pages} {929} (\bibinfo {year} {1990})}\BibitemShut {NoStop}%
\bibitem [{\citenamefont {Mahan}(2000)}]{Mahan_2000}%
  \BibitemOpen
  \bibfield  {author} {\bibinfo {author} {\bibfnamefont {G.~D.}\ \bibnamefont {Mahan}},\ }\href {https://doi.org/10.1007/978-1-4757-5714-9} {\emph {\bibinfo {title} {Many-Particle Physics}}}\ (\bibinfo  {publisher} {Springer US},\ \bibinfo {year} {2000})\BibitemShut {NoStop}%
\bibitem [{\citenamefont {Lee}\ \emph {et~al.}(1987)\citenamefont {Lee}, \citenamefont {Iwasa},\ and\ \citenamefont {Miura}}]{Lee_1987}%
  \BibitemOpen
  \bibfield  {author} {\bibinfo {author} {\bibfnamefont {J.~S.}\ \bibnamefont {Lee}}, \bibinfo {author} {\bibfnamefont {Y.}~\bibnamefont {Iwasa}},\ and\ \bibinfo {author} {\bibfnamefont {N.}~\bibnamefont {Miura}},\ }\bibfield  {title} {\bibinfo {title} {Observation of the fermi edge anomaly in the absorption and luminescence spectra of n-type modulation-doped gaas-algaas quantum wells},\ }\href {https://doi.org/10.1088/0268-1242/2/10/008} {\bibfield  {journal} {\bibinfo  {journal} {Semiconductor Science and Technology}\ }\textbf {\bibinfo {volume} {2}},\ \bibinfo {pages} {675} (\bibinfo {year} {1987})}\BibitemShut {NoStop}%
\bibitem [{\citenamefont {Skolnick}\ \emph {et~al.}(1987)\citenamefont {Skolnick}, \citenamefont {Rorison}, \citenamefont {Nash}, \citenamefont {Mowbray}, \citenamefont {Tapster}, \citenamefont {Bass},\ and\ \citenamefont {Pitt}}]{Skolnick_1987}%
  \BibitemOpen
  \bibfield  {author} {\bibinfo {author} {\bibfnamefont {M.~S.}\ \bibnamefont {Skolnick}}, \bibinfo {author} {\bibfnamefont {J.~M.}\ \bibnamefont {Rorison}}, \bibinfo {author} {\bibfnamefont {K.~J.}\ \bibnamefont {Nash}}, \bibinfo {author} {\bibfnamefont {D.~J.}\ \bibnamefont {Mowbray}}, \bibinfo {author} {\bibfnamefont {P.~R.}\ \bibnamefont {Tapster}}, \bibinfo {author} {\bibfnamefont {S.~J.}\ \bibnamefont {Bass}},\ and\ \bibinfo {author} {\bibfnamefont {A.~D.}\ \bibnamefont {Pitt}},\ }\bibfield  {title} {\bibinfo {title} {Observation of a many-body edge singularity in quantum-well luminescence spectra},\ }\href {https://doi.org/10.1103/physrevlett.58.2130} {\bibfield  {journal} {\bibinfo  {journal} {Physical Review Letters}\ }\textbf {\bibinfo {volume} {58}},\ \bibinfo {pages} {2130} (\bibinfo {year} {1987})}\BibitemShut {NoStop}%
\bibitem [{\citenamefont {Calleja}\ \emph {et~al.}(1991)\citenamefont {Calleja}, \citenamefont {Goñi}, \citenamefont {Dennis}, \citenamefont {Weiner}, \citenamefont {Pinczuk}, \citenamefont {Schmitt-Rink}, \citenamefont {Pfeiffer}, \citenamefont {West}, \citenamefont {Müller},\ and\ \citenamefont {Ruckenstein}}]{Calleja_1991}%
  \BibitemOpen
  \bibfield  {author} {\bibinfo {author} {\bibfnamefont {J.}~\bibnamefont {Calleja}}, \bibinfo {author} {\bibfnamefont {A.}~\bibnamefont {Goñi}}, \bibinfo {author} {\bibfnamefont {B.}~\bibnamefont {Dennis}}, \bibinfo {author} {\bibfnamefont {J.}~\bibnamefont {Weiner}}, \bibinfo {author} {\bibfnamefont {A.}~\bibnamefont {Pinczuk}}, \bibinfo {author} {\bibfnamefont {S.}~\bibnamefont {Schmitt-Rink}}, \bibinfo {author} {\bibfnamefont {L.}~\bibnamefont {Pfeiffer}}, \bibinfo {author} {\bibfnamefont {K.}~\bibnamefont {West}}, \bibinfo {author} {\bibfnamefont {J.}~\bibnamefont {Müller}},\ and\ \bibinfo {author} {\bibfnamefont {A.}~\bibnamefont {Ruckenstein}},\ }\bibfield  {title} {\bibinfo {title} {Large optical singularities of the one-dimensional electron gas in semiconductor quantum wires},\ }\href {https://doi.org/10.1016/0038-1098(91)90442-x} {\bibfield  {journal} {\bibinfo  {journal} {Solid State Communications}\ }\textbf {\bibinfo {volume} {79}},\ \bibinfo {pages} {911} (\bibinfo {year} {1991})}\BibitemShut
  {NoStop}%
\bibitem [{\citenamefont {Hentschel}\ \emph {et~al.}(2005)\citenamefont {Hentschel}, \citenamefont {Ullmo},\ and\ \citenamefont {Baranger}}]{Hentschel_2005}%
  \BibitemOpen
  \bibfield  {author} {\bibinfo {author} {\bibfnamefont {M.}~\bibnamefont {Hentschel}}, \bibinfo {author} {\bibfnamefont {D.}~\bibnamefont {Ullmo}},\ and\ \bibinfo {author} {\bibfnamefont {H.~U.}\ \bibnamefont {Baranger}},\ }\bibfield  {title} {\bibinfo {title} {Fermi edge singularities in the mesoscopic regime: Anderson orthogonality catastrophe},\ }\href {https://doi.org/10.1103/physrevb.72.035310} {\bibfield  {journal} {\bibinfo  {journal} {Physical Review B}\ }\textbf {\bibinfo {volume} {72}},\ \bibinfo {pages} {035310} (\bibinfo {year} {2005})}\BibitemShut {NoStop}%
\bibitem [{\citenamefont {Hentschel}\ \emph {et~al.}(2007)\citenamefont {Hentschel}, \citenamefont {Ullmo},\ and\ \citenamefont {Baranger}}]{Hentschel_2007}%
  \BibitemOpen
  \bibfield  {author} {\bibinfo {author} {\bibfnamefont {M.}~\bibnamefont {Hentschel}}, \bibinfo {author} {\bibfnamefont {D.}~\bibnamefont {Ullmo}},\ and\ \bibinfo {author} {\bibfnamefont {H.~U.}\ \bibnamefont {Baranger}},\ }\bibfield  {title} {\bibinfo {title} {Fermi edge singularities in the mesoscopic regime: Photoabsorption spectra},\ }\href {https://doi.org/10.1103/physrevb.76.245419} {\bibfield  {journal} {\bibinfo  {journal} {Physical Review B}\ }\textbf {\bibinfo {volume} {76}},\ \bibinfo {pages} {245419} (\bibinfo {year} {2007})}\BibitemShut {NoStop}%
\bibitem [{\citenamefont {Heyl}\ and\ \citenamefont {Kehrein}(2012)}]{Heyl_2012}%
  \BibitemOpen
  \bibfield  {author} {\bibinfo {author} {\bibfnamefont {M.}~\bibnamefont {Heyl}}\ and\ \bibinfo {author} {\bibfnamefont {S.}~\bibnamefont {Kehrein}},\ }\bibfield  {title} {\bibinfo {title} {X-ray edge singularity in optical spectra of quantum dots},\ }\href {https://doi.org/10.1103/physrevb.85.155413} {\bibfield  {journal} {\bibinfo  {journal} {Physical Review B}\ }\textbf {\bibinfo {volume} {85}},\ \bibinfo {pages} {155413} (\bibinfo {year} {2012})}\BibitemShut {NoStop}%
\bibitem [{\citenamefont {Goold}\ \emph {et~al.}(2011)\citenamefont {Goold}, \citenamefont {Fogarty}, \citenamefont {Lo~Gullo}, \citenamefont {Paternostro},\ and\ \citenamefont {Busch}}]{Goold_2011}%
  \BibitemOpen
  \bibfield  {author} {\bibinfo {author} {\bibfnamefont {J.}~\bibnamefont {Goold}}, \bibinfo {author} {\bibfnamefont {T.}~\bibnamefont {Fogarty}}, \bibinfo {author} {\bibfnamefont {N.}~\bibnamefont {Lo~Gullo}}, \bibinfo {author} {\bibfnamefont {M.}~\bibnamefont {Paternostro}},\ and\ \bibinfo {author} {\bibfnamefont {T.}~\bibnamefont {Busch}},\ }\bibfield  {title} {\bibinfo {title} {Orthogonality catastrophe as a consequence of qubit embedding in an ultracold fermi gas},\ }\href {https://doi.org/10.1103/physreva.84.063632} {\bibfield  {journal} {\bibinfo  {journal} {Physical Review A}\ }\textbf {\bibinfo {volume} {84}},\ \bibinfo {pages} {063632} (\bibinfo {year} {2011})}\BibitemShut {NoStop}%
\bibitem [{\citenamefont {Knap}\ \emph {et~al.}(2012)\citenamefont {Knap}, \citenamefont {Shashi}, \citenamefont {Nishida}, \citenamefont {Imambekov}, \citenamefont {Abanin},\ and\ \citenamefont {Demler}}]{Knap_2012}%
  \BibitemOpen
  \bibfield  {author} {\bibinfo {author} {\bibfnamefont {M.}~\bibnamefont {Knap}}, \bibinfo {author} {\bibfnamefont {A.}~\bibnamefont {Shashi}}, \bibinfo {author} {\bibfnamefont {Y.}~\bibnamefont {Nishida}}, \bibinfo {author} {\bibfnamefont {A.}~\bibnamefont {Imambekov}}, \bibinfo {author} {\bibfnamefont {D.~A.}\ \bibnamefont {Abanin}},\ and\ \bibinfo {author} {\bibfnamefont {E.}~\bibnamefont {Demler}},\ }\bibfield  {title} {\bibinfo {title} {Time-dependent impurity in ultracold fermions: Orthogonality catastrophe and beyond},\ }\href {https://doi.org/10.1103/physrevx.2.041020} {\bibfield  {journal} {\bibinfo  {journal} {Physical Review X}\ }\textbf {\bibinfo {volume} {2}},\ \bibinfo {pages} {041020} (\bibinfo {year} {2012})}\BibitemShut {NoStop}%
\bibitem [{\citenamefont {Sindona}\ \emph {et~al.}(2013)\citenamefont {Sindona}, \citenamefont {Goold}, \citenamefont {Lo~Gullo}, \citenamefont {Lorenzo},\ and\ \citenamefont {Plastina}}]{Sindona_2013}%
  \BibitemOpen
  \bibfield  {author} {\bibinfo {author} {\bibfnamefont {A.}~\bibnamefont {Sindona}}, \bibinfo {author} {\bibfnamefont {J.}~\bibnamefont {Goold}}, \bibinfo {author} {\bibfnamefont {N.}~\bibnamefont {Lo~Gullo}}, \bibinfo {author} {\bibfnamefont {S.}~\bibnamefont {Lorenzo}},\ and\ \bibinfo {author} {\bibfnamefont {F.}~\bibnamefont {Plastina}},\ }\bibfield  {title} {\bibinfo {title} {Orthogonality catastrophe and decoherence in a trapped-fermion environment},\ }\href {https://doi.org/10.1103/physrevlett.111.165303} {\bibfield  {journal} {\bibinfo  {journal} {Physical Review Letters}\ }\textbf {\bibinfo {volume} {111}},\ \bibinfo {pages} {165303} (\bibinfo {year} {2013})}\BibitemShut {NoStop}%
\bibitem [{\citenamefont {Cetina}\ \emph {et~al.}(2016)\citenamefont {Cetina}, \citenamefont {Jag}, \citenamefont {Lous}, \citenamefont {Fritsche}, \citenamefont {Walraven}, \citenamefont {Grimm}, \citenamefont {Levinsen}, \citenamefont {Parish}, \citenamefont {Schmidt}, \citenamefont {Knap},\ and\ \citenamefont {Demler}}]{Cetina_2016}%
  \BibitemOpen
  \bibfield  {author} {\bibinfo {author} {\bibfnamefont {M.}~\bibnamefont {Cetina}}, \bibinfo {author} {\bibfnamefont {M.}~\bibnamefont {Jag}}, \bibinfo {author} {\bibfnamefont {R.~S.}\ \bibnamefont {Lous}}, \bibinfo {author} {\bibfnamefont {I.}~\bibnamefont {Fritsche}}, \bibinfo {author} {\bibfnamefont {J.~T.~M.}\ \bibnamefont {Walraven}}, \bibinfo {author} {\bibfnamefont {R.}~\bibnamefont {Grimm}}, \bibinfo {author} {\bibfnamefont {J.}~\bibnamefont {Levinsen}}, \bibinfo {author} {\bibfnamefont {M.~M.}\ \bibnamefont {Parish}}, \bibinfo {author} {\bibfnamefont {R.}~\bibnamefont {Schmidt}}, \bibinfo {author} {\bibfnamefont {M.}~\bibnamefont {Knap}},\ and\ \bibinfo {author} {\bibfnamefont {E.}~\bibnamefont {Demler}},\ }\bibfield  {title} {\bibinfo {title} {Ultrafast many-body interferometry of impurities coupled to a fermi sea},\ }\href {https://doi.org/10.1126/science.aaf5134} {\bibfield  {journal} {\bibinfo  {journal} {Science}\ }\textbf {\bibinfo {volume} {354}},\ \bibinfo {pages} {96} (\bibinfo {year}
  {2016})}\BibitemShut {NoStop}%
\bibitem [{\citenamefont {Schmidt}\ \emph {et~al.}(2018)\citenamefont {Schmidt}, \citenamefont {Knap}, \citenamefont {Ivanov}, \citenamefont {You}, \citenamefont {Cetina},\ and\ \citenamefont {Demler}}]{Schmidt_2018}%
  \BibitemOpen
  \bibfield  {author} {\bibinfo {author} {\bibfnamefont {R.}~\bibnamefont {Schmidt}}, \bibinfo {author} {\bibfnamefont {M.}~\bibnamefont {Knap}}, \bibinfo {author} {\bibfnamefont {D.~A.}\ \bibnamefont {Ivanov}}, \bibinfo {author} {\bibfnamefont {J.-S.}\ \bibnamefont {You}}, \bibinfo {author} {\bibfnamefont {M.}~\bibnamefont {Cetina}},\ and\ \bibinfo {author} {\bibfnamefont {E.}~\bibnamefont {Demler}},\ }\bibfield  {title} {\bibinfo {title} {Universal many-body response of heavy impurities coupled to a fermi sea: a review of recent progress},\ }\href {https://doi.org/10.1088/1361-6633/aa9593} {\bibfield  {journal} {\bibinfo  {journal} {Reports on Progress in Physics}\ }\textbf {\bibinfo {volume} {81}},\ \bibinfo {pages} {024401} (\bibinfo {year} {2018})}\BibitemShut {NoStop}%
\bibitem [{\citenamefont {Matveev}\ and\ \citenamefont {Larkin}(1992)}]{Matveev_1992}%
  \BibitemOpen
  \bibfield  {author} {\bibinfo {author} {\bibfnamefont {K.~A.}\ \bibnamefont {Matveev}}\ and\ \bibinfo {author} {\bibfnamefont {A.~I.}\ \bibnamefont {Larkin}},\ }\bibfield  {title} {\bibinfo {title} {Interaction-induced threshold singularities in tunneling via localized levels},\ }\href {https://doi.org/10.1103/physrevb.46.15337} {\bibfield  {journal} {\bibinfo  {journal} {Physical Review B}\ }\textbf {\bibinfo {volume} {46}},\ \bibinfo {pages} {15337} (\bibinfo {year} {1992})}\BibitemShut {NoStop}%
\bibitem [{\citenamefont {Geim}\ \emph {et~al.}(1994)\citenamefont {Geim}, \citenamefont {Main}, \citenamefont {La~Scala}, \citenamefont {Eaves}, \citenamefont {Foster}, \citenamefont {Beton}, \citenamefont {Sakai}, \citenamefont {Sheard}, \citenamefont {Henini}, \citenamefont {Hill},\ and\ \citenamefont {Pate}}]{Geim_1994}%
  \BibitemOpen
  \bibfield  {author} {\bibinfo {author} {\bibfnamefont {A.~K.}\ \bibnamefont {Geim}}, \bibinfo {author} {\bibfnamefont {P.~C.}\ \bibnamefont {Main}}, \bibinfo {author} {\bibfnamefont {N.}~\bibnamefont {La~Scala}}, \bibinfo {author} {\bibfnamefont {L.}~\bibnamefont {Eaves}}, \bibinfo {author} {\bibfnamefont {T.~J.}\ \bibnamefont {Foster}}, \bibinfo {author} {\bibfnamefont {P.~H.}\ \bibnamefont {Beton}}, \bibinfo {author} {\bibfnamefont {J.~W.}\ \bibnamefont {Sakai}}, \bibinfo {author} {\bibfnamefont {F.~W.}\ \bibnamefont {Sheard}}, \bibinfo {author} {\bibfnamefont {M.}~\bibnamefont {Henini}}, \bibinfo {author} {\bibfnamefont {G.}~\bibnamefont {Hill}},\ and\ \bibinfo {author} {\bibfnamefont {M.~A.}\ \bibnamefont {Pate}},\ }\bibfield  {title} {\bibinfo {title} {Fermi-edge singularity in resonant tunneling},\ }\href {https://doi.org/10.1103/physrevlett.72.2061} {\bibfield  {journal} {\bibinfo  {journal} {Physical Review Letters}\ }\textbf {\bibinfo {volume} {72}},\ \bibinfo {pages} {2061} (\bibinfo {year}
  {1994})}\BibitemShut {NoStop}%
\bibitem [{\citenamefont {Cobden}\ and\ \citenamefont {Muzykantskii}(1995)}]{Cobden_1995}%
  \BibitemOpen
  \bibfield  {author} {\bibinfo {author} {\bibfnamefont {D.~H.}\ \bibnamefont {Cobden}}\ and\ \bibinfo {author} {\bibfnamefont {B.~A.}\ \bibnamefont {Muzykantskii}},\ }\bibfield  {title} {\bibinfo {title} {Finite-temperature fermi-edge singularity in tunneling studied using random telegraph signals},\ }\href {https://doi.org/10.1103/physrevlett.75.4274} {\bibfield  {journal} {\bibinfo  {journal} {Physical Review Letters}\ }\textbf {\bibinfo {volume} {75}},\ \bibinfo {pages} {4274} (\bibinfo {year} {1995})}\BibitemShut {NoStop}%
\bibitem [{\citenamefont {Benedict}\ \emph {et~al.}(1998)\citenamefont {Benedict}, \citenamefont {Thornton}, \citenamefont {Ihn}, \citenamefont {Main}, \citenamefont {Eaves},\ and\ \citenamefont {Henini}}]{Benedict_1998}%
  \BibitemOpen
  \bibfield  {author} {\bibinfo {author} {\bibfnamefont {K.}~\bibnamefont {Benedict}}, \bibinfo {author} {\bibfnamefont {A.}~\bibnamefont {Thornton}}, \bibinfo {author} {\bibfnamefont {T.}~\bibnamefont {Ihn}}, \bibinfo {author} {\bibfnamefont {P.}~\bibnamefont {Main}}, \bibinfo {author} {\bibfnamefont {L.}~\bibnamefont {Eaves}},\ and\ \bibinfo {author} {\bibfnamefont {M.}~\bibnamefont {Henini}},\ }\bibfield  {title} {\bibinfo {title} {Fermi edge singularities in high magnetic fields},\ }\href {https://doi.org/10.1016/s0921-4526(98)00562-6} {\bibfield  {journal} {\bibinfo  {journal} {Physica B: Condensed Matter}\ }\textbf {\bibinfo {volume} {256}},\ \bibinfo {pages} {519} (\bibinfo {year} {1998})}\BibitemShut {NoStop}%
\bibitem [{\citenamefont {Hapke-Wurst}\ \emph {et~al.}(2000)\citenamefont {Hapke-Wurst}, \citenamefont {Zeitler}, \citenamefont {Frahm}, \citenamefont {Jansen}, \citenamefont {Haug},\ and\ \citenamefont {Pierz}}]{Hapke_Wurst_2000}%
  \BibitemOpen
  \bibfield  {author} {\bibinfo {author} {\bibfnamefont {I.}~\bibnamefont {Hapke-Wurst}}, \bibinfo {author} {\bibfnamefont {U.}~\bibnamefont {Zeitler}}, \bibinfo {author} {\bibfnamefont {H.}~\bibnamefont {Frahm}}, \bibinfo {author} {\bibfnamefont {A.~G.~M.}\ \bibnamefont {Jansen}}, \bibinfo {author} {\bibfnamefont {R.~J.}\ \bibnamefont {Haug}},\ and\ \bibinfo {author} {\bibfnamefont {K.}~\bibnamefont {Pierz}},\ }\bibfield  {title} {\bibinfo {title} {Magnetic-field-induced singularities in spin-dependent tunneling through inas quantum dots},\ }\href {https://doi.org/10.1103/physrevb.62.12621} {\bibfield  {journal} {\bibinfo  {journal} {Physical Review B}\ }\textbf {\bibinfo {volume} {62}},\ \bibinfo {pages} {12621} (\bibinfo {year} {2000})}\BibitemShut {NoStop}%
\bibitem [{\citenamefont {Frahm}\ \emph {et~al.}(2006)\citenamefont {Frahm}, \citenamefont {von Zobeltitz}, \citenamefont {Maire},\ and\ \citenamefont {Haug}}]{Frahm_2006}%
  \BibitemOpen
  \bibfield  {author} {\bibinfo {author} {\bibfnamefont {H.}~\bibnamefont {Frahm}}, \bibinfo {author} {\bibfnamefont {C.}~\bibnamefont {von Zobeltitz}}, \bibinfo {author} {\bibfnamefont {N.}~\bibnamefont {Maire}},\ and\ \bibinfo {author} {\bibfnamefont {R.~J.}\ \bibnamefont {Haug}},\ }\bibfield  {title} {\bibinfo {title} {Fermi-edge singularities in transport through quantum dots},\ }\href {https://doi.org/10.1103/physrevb.74.035329} {\bibfield  {journal} {\bibinfo  {journal} {Physical Review B}\ }\textbf {\bibinfo {volume} {74}},\ \bibinfo {pages} {035329} (\bibinfo {year} {2006})}\BibitemShut {NoStop}%
\bibitem [{\citenamefont {Ubbelohde}\ \emph {et~al.}(2012)\citenamefont {Ubbelohde}, \citenamefont {Roszak}, \citenamefont {Hohls}, \citenamefont {Maire}, \citenamefont {Haug},\ and\ \citenamefont {Novotný}}]{Ubbelohde_2012}%
  \BibitemOpen
  \bibfield  {author} {\bibinfo {author} {\bibfnamefont {N.}~\bibnamefont {Ubbelohde}}, \bibinfo {author} {\bibfnamefont {K.}~\bibnamefont {Roszak}}, \bibinfo {author} {\bibfnamefont {F.}~\bibnamefont {Hohls}}, \bibinfo {author} {\bibfnamefont {N.}~\bibnamefont {Maire}}, \bibinfo {author} {\bibfnamefont {R.~J.}\ \bibnamefont {Haug}},\ and\ \bibinfo {author} {\bibfnamefont {T.}~\bibnamefont {Novotný}},\ }\bibfield  {title} {\bibinfo {title} {Strong quantum memory at resonant fermi edges revealed by shot noise},\ }\href {https://doi.org/10.1038/srep00374} {\bibfield  {journal} {\bibinfo  {journal} {Scientific Reports}\ }\textbf {\bibinfo {volume} {2}},\ \bibinfo {pages} {374} (\bibinfo {year} {2012})}\BibitemShut {NoStop}%
\bibitem [{\citenamefont {Keeling}\ \emph {et~al.}(2006)\citenamefont {Keeling}, \citenamefont {Klich},\ and\ \citenamefont {Levitov}}]{keeling_2006_minim}%
  \BibitemOpen
  \bibfield  {author} {\bibinfo {author} {\bibfnamefont {J.}~\bibnamefont {Keeling}}, \bibinfo {author} {\bibfnamefont {I.}~\bibnamefont {Klich}},\ and\ \bibinfo {author} {\bibfnamefont {L.~S.}\ \bibnamefont {Levitov}},\ }\bibfield  {title} {\bibinfo {title} {Minimal excitation states of electrons in one-dimensional wires},\ }\href {https://doi.org/10.1103/physrevlett.97.116403} {\bibfield  {journal} {\bibinfo  {journal} {Physical Review Letters}\ }\textbf {\bibinfo {volume} {97}},\ \bibinfo {pages} {116403} (\bibinfo {year} {2006})}\BibitemShut {NoStop}%
\bibitem [{\citenamefont {Dubois}\ \emph {et~al.}(2013{\natexlab{a}})\citenamefont {Dubois}, \citenamefont {Jullien}, \citenamefont {Portier}, \citenamefont {Roche}, \citenamefont {Cavanna}, \citenamefont {Jin}, \citenamefont {Wegscheider}, \citenamefont {Roulleau},\ and\ \citenamefont {Glattli}}]{dubois_2013_minim}%
  \BibitemOpen
  \bibfield  {author} {\bibinfo {author} {\bibfnamefont {J.}~\bibnamefont {Dubois}}, \bibinfo {author} {\bibfnamefont {T.}~\bibnamefont {Jullien}}, \bibinfo {author} {\bibfnamefont {F.}~\bibnamefont {Portier}}, \bibinfo {author} {\bibfnamefont {P.}~\bibnamefont {Roche}}, \bibinfo {author} {\bibfnamefont {A.}~\bibnamefont {Cavanna}}, \bibinfo {author} {\bibfnamefont {Y.}~\bibnamefont {Jin}}, \bibinfo {author} {\bibfnamefont {W.}~\bibnamefont {Wegscheider}}, \bibinfo {author} {\bibfnamefont {P.}~\bibnamefont {Roulleau}},\ and\ \bibinfo {author} {\bibfnamefont {D.~C.}\ \bibnamefont {Glattli}},\ }\bibfield  {title} {\bibinfo {title} {Minimal-excitation states for electron quantum optics using levitons},\ }\href {https://doi.org/10.1038/nature12713} {\bibfield  {journal} {\bibinfo  {journal} {Nature}\ }\textbf {\bibinfo {volume} {502}},\ \bibinfo {pages} {659} (\bibinfo {year} {2013}{\natexlab{a}})}\BibitemShut {NoStop}%
\bibitem [{\citenamefont {Gabelli}\ and\ \citenamefont {Reulet}(2013)}]{gabelli_2013_shapin}%
  \BibitemOpen
  \bibfield  {author} {\bibinfo {author} {\bibfnamefont {J.}~\bibnamefont {Gabelli}}\ and\ \bibinfo {author} {\bibfnamefont {B.}~\bibnamefont {Reulet}},\ }\bibfield  {title} {\bibinfo {title} {Shaping a time-dependent excitation to minimize the shot noise in a tunnel junction},\ }\href {https://doi.org/10.1103/physrevb.87.075403} {\bibfield  {journal} {\bibinfo  {journal} {Physical Review B}\ }\textbf {\bibinfo {volume} {87}},\ \bibinfo {pages} {075403} (\bibinfo {year} {2013})}\BibitemShut {NoStop}%
\bibitem [{\citenamefont {Gaury}\ and\ \citenamefont {Waintal}(2014)}]{Gaury_2014}%
  \BibitemOpen
  \bibfield  {author} {\bibinfo {author} {\bibfnamefont {B.}~\bibnamefont {Gaury}}\ and\ \bibinfo {author} {\bibfnamefont {X.}~\bibnamefont {Waintal}},\ }\bibfield  {title} {\bibinfo {title} {Dynamical control of interference using voltage pulses in the quantum regime},\ }\href {https://doi.org/10.1038/ncomms4844} {\bibfield  {journal} {\bibinfo  {journal} {Nature Communications}\ }\textbf {\bibinfo {volume} {5}},\ \bibinfo {pages} {3844} (\bibinfo {year} {2014})}\BibitemShut {NoStop}%
\bibitem [{\citenamefont {Glattli}\ and\ \citenamefont {Roulleau}(2016)}]{glattli_2016_levit}%
  \BibitemOpen
  \bibfield  {author} {\bibinfo {author} {\bibfnamefont {D.~C.}\ \bibnamefont {Glattli}}\ and\ \bibinfo {author} {\bibfnamefont {P.~S.}\ \bibnamefont {Roulleau}},\ }\bibfield  {title} {\bibinfo {title} {Levitons for electron quantum optics},\ }\href {https://doi.org/10.1002/pssb.201600650} {\bibfield  {journal} {\bibinfo  {journal} {physica status solidi (b)}\ }\textbf {\bibinfo {volume} {254}},\ \bibinfo {pages} {1} (\bibinfo {year} {2016})}\BibitemShut {NoStop}%
\bibitem [{\citenamefont {B{\"a}uerle}\ \emph {et~al.}(2018)\citenamefont {B{\"a}uerle}, \citenamefont {Glattli}, \citenamefont {Meunier}, \citenamefont {Portier}, \citenamefont {Roche}, \citenamefont {Roulleau}, \citenamefont {Takada},\ and\ \citenamefont {Waintal}}]{baeuerle_2018_coher}%
  \BibitemOpen
  \bibfield  {author} {\bibinfo {author} {\bibfnamefont {C.}~\bibnamefont {B{\"a}uerle}}, \bibinfo {author} {\bibfnamefont {D.~C.}\ \bibnamefont {Glattli}}, \bibinfo {author} {\bibfnamefont {T.}~\bibnamefont {Meunier}}, \bibinfo {author} {\bibfnamefont {F.}~\bibnamefont {Portier}}, \bibinfo {author} {\bibfnamefont {P.}~\bibnamefont {Roche}}, \bibinfo {author} {\bibfnamefont {P.}~\bibnamefont {Roulleau}}, \bibinfo {author} {\bibfnamefont {S.}~\bibnamefont {Takada}},\ and\ \bibinfo {author} {\bibfnamefont {X.}~\bibnamefont {Waintal}},\ }\bibfield  {title} {\bibinfo {title} {Coherent control of single electrons: a review of current progress},\ }\href {https://doi.org/10.1088/1361-6633/aaa98a} {\bibfield  {journal} {\bibinfo  {journal} {Reports on Progress in Physics}\ }\textbf {\bibinfo {volume} {81}},\ \bibinfo {pages} {056503} (\bibinfo {year} {2018})}\BibitemShut {NoStop}%
\bibitem [{\citenamefont {Ivanov}\ \emph {et~al.}(1995)\citenamefont {Ivanov}, \citenamefont {Lee},\ and\ \citenamefont {Levitov}}]{ivanov_1995_coher}%
  \BibitemOpen
  \bibfield  {author} {\bibinfo {author} {\bibfnamefont {D.~A.}\ \bibnamefont {Ivanov}}, \bibinfo {author} {\bibfnamefont {H.~W.}\ \bibnamefont {Lee}},\ and\ \bibinfo {author} {\bibfnamefont {L.~S.}\ \bibnamefont {Levitov}},\ }\bibfield  {title} {\bibinfo {title} {Coherent states of alternating current},\ }\href {https://doi.org/10.1103/PhysRevB.56.6839} {\bibfield  {journal} {\bibinfo  {journal} {Physical Review B}\ }\textbf {\bibinfo {volume} {56}},\ \bibinfo {pages} {6839} (\bibinfo {year} {1995})}\BibitemShut {NoStop}%
\bibitem [{\citenamefont {Levitov}\ \emph {et~al.}(1996)\citenamefont {Levitov}, \citenamefont {Lee},\ and\ \citenamefont {Lesovik}}]{levitov_1996_elect}%
  \BibitemOpen
  \bibfield  {author} {\bibinfo {author} {\bibfnamefont {L.~S.}\ \bibnamefont {Levitov}}, \bibinfo {author} {\bibfnamefont {H.}~\bibnamefont {Lee}},\ and\ \bibinfo {author} {\bibfnamefont {G.~B.}\ \bibnamefont {Lesovik}},\ }\bibfield  {title} {\bibinfo {title} {Electron counting statistics and coherent states of electric current},\ }\href {https://doi.org/10.1063/1.531672} {\bibfield  {journal} {\bibinfo  {journal} {Journal of Mathematical Physics}\ }\textbf {\bibinfo {volume} {37}},\ \bibinfo {pages} {4845} (\bibinfo {year} {1996})}\BibitemShut {NoStop}%
\bibitem [{\citenamefont {Dubois}\ \emph {et~al.}(2013{\natexlab{b}})\citenamefont {Dubois}, \citenamefont {Jullien}, \citenamefont {Grenier}, \citenamefont {Degiovanni}, \citenamefont {Roulleau},\ and\ \citenamefont {Glattli}}]{dubois_2013_integ}%
  \BibitemOpen
  \bibfield  {author} {\bibinfo {author} {\bibfnamefont {J.}~\bibnamefont {Dubois}}, \bibinfo {author} {\bibfnamefont {T.}~\bibnamefont {Jullien}}, \bibinfo {author} {\bibfnamefont {C.}~\bibnamefont {Grenier}}, \bibinfo {author} {\bibfnamefont {P.}~\bibnamefont {Degiovanni}}, \bibinfo {author} {\bibfnamefont {P.}~\bibnamefont {Roulleau}},\ and\ \bibinfo {author} {\bibfnamefont {D.~C.}\ \bibnamefont {Glattli}},\ }\bibfield  {title} {\bibinfo {title} {Integer and fractional charge lorentzian voltage pulses analyzed in the framework of photon-assisted shot noise},\ }\href {https://doi.org/10.1103/physrevb.88.085301} {\bibfield  {journal} {\bibinfo  {journal} {Physical Review B}\ }\textbf {\bibinfo {volume} {88}},\ \bibinfo {pages} {085301} (\bibinfo {year} {2013}{\natexlab{b}})}\BibitemShut {NoStop}%
\bibitem [{\citenamefont {Hofer}\ and\ \citenamefont {Flindt}(2014)}]{hofer_2014_mach}%
  \BibitemOpen
  \bibfield  {author} {\bibinfo {author} {\bibfnamefont {P.~P.}\ \bibnamefont {Hofer}}\ and\ \bibinfo {author} {\bibfnamefont {C.}~\bibnamefont {Flindt}},\ }\bibfield  {title} {\bibinfo {title} {Mach-zehnder interferometry with periodic voltage pulses},\ }\href {https://doi.org/10.1103/physrevb.90.235416} {\bibfield  {journal} {\bibinfo  {journal} {Physical Review B}\ }\textbf {\bibinfo {volume} {90}},\ \bibinfo {pages} {235416} (\bibinfo {year} {2014})}\BibitemShut {NoStop}%
\bibitem [{\citenamefont {Glattli}\ and\ \citenamefont {Roulleau}(2017)}]{glattli_2017_pseud}%
  \BibitemOpen
  \bibfield  {author} {\bibinfo {author} {\bibfnamefont {D.~C.}\ \bibnamefont {Glattli}}\ and\ \bibinfo {author} {\bibfnamefont {P.}~\bibnamefont {Roulleau}},\ }\bibfield  {title} {\bibinfo {title} {Pseudorandom binary injection of levitons for electron quantum optics},\ }\href {https://doi.org/10.1103/PhysRevB.97.125407} {\bibfield  {journal} {\bibinfo  {journal} {Physical Review B}\ }\textbf {\bibinfo {volume} {97}},\ \bibinfo {pages} {125407} (\bibinfo {year} {2017})}\BibitemShut {NoStop}%
\bibitem [{\citenamefont {Jullien}\ \emph {et~al.}(2014)\citenamefont {Jullien}, \citenamefont {Roulleau}, \citenamefont {Roche}, \citenamefont {Cavanna}, \citenamefont {Jin},\ and\ \citenamefont {Glattli}}]{jullien_2014_quant}%
  \BibitemOpen
  \bibfield  {author} {\bibinfo {author} {\bibfnamefont {T.}~\bibnamefont {Jullien}}, \bibinfo {author} {\bibfnamefont {P.}~\bibnamefont {Roulleau}}, \bibinfo {author} {\bibfnamefont {B.}~\bibnamefont {Roche}}, \bibinfo {author} {\bibfnamefont {A.}~\bibnamefont {Cavanna}}, \bibinfo {author} {\bibfnamefont {Y.}~\bibnamefont {Jin}},\ and\ \bibinfo {author} {\bibfnamefont {D.~C.}\ \bibnamefont {Glattli}},\ }\bibfield  {title} {\bibinfo {title} {Quantum tomography of an electron},\ }\href {https://doi.org/10.1038/nature13821} {\bibfield  {journal} {\bibinfo  {journal} {Nature}\ }\textbf {\bibinfo {volume} {514}},\ \bibinfo {pages} {603} (\bibinfo {year} {2014})}\BibitemShut {NoStop}%
\bibitem [{\citenamefont {Bisognin}\ \emph {et~al.}(2019)\citenamefont {Bisognin}, \citenamefont {Marguerite}, \citenamefont {Roussel}, \citenamefont {Kumar}, \citenamefont {Cabart}, \citenamefont {Chapdelaine}, \citenamefont {Mohammad-Djafari}, \citenamefont {Berroir}, \citenamefont {Bocquillon}, \citenamefont {Pla{\c{c}}ais}, \citenamefont {Cavanna}, \citenamefont {Gennser}, \citenamefont {Jin}, \citenamefont {Degiovanni},\ and\ \citenamefont {F{\`e}ve}}]{bisognin_2019_quant}%
  \BibitemOpen
  \bibfield  {author} {\bibinfo {author} {\bibfnamefont {R.}~\bibnamefont {Bisognin}}, \bibinfo {author} {\bibfnamefont {A.}~\bibnamefont {Marguerite}}, \bibinfo {author} {\bibfnamefont {B.}~\bibnamefont {Roussel}}, \bibinfo {author} {\bibfnamefont {M.}~\bibnamefont {Kumar}}, \bibinfo {author} {\bibfnamefont {C.}~\bibnamefont {Cabart}}, \bibinfo {author} {\bibfnamefont {C.}~\bibnamefont {Chapdelaine}}, \bibinfo {author} {\bibfnamefont {A.}~\bibnamefont {Mohammad-Djafari}}, \bibinfo {author} {\bibfnamefont {J.-M.}\ \bibnamefont {Berroir}}, \bibinfo {author} {\bibfnamefont {E.}~\bibnamefont {Bocquillon}}, \bibinfo {author} {\bibfnamefont {B.}~\bibnamefont {Pla{\c{c}}ais}}, \bibinfo {author} {\bibfnamefont {A.}~\bibnamefont {Cavanna}}, \bibinfo {author} {\bibfnamefont {U.}~\bibnamefont {Gennser}}, \bibinfo {author} {\bibfnamefont {Y.}~\bibnamefont {Jin}}, \bibinfo {author} {\bibfnamefont {P.}~\bibnamefont {Degiovanni}},\ and\ \bibinfo {author} {\bibfnamefont {G.}~\bibnamefont {F{\`e}ve}},\ }\bibfield  {title}
  {\bibinfo {title} {Quantum tomography of electrical currents},\ }\href {https://doi.org/10.1038/s41467-019-11369-5} {\bibfield  {journal} {\bibinfo  {journal} {Nature Communications}\ }\textbf {\bibinfo {volume} {10}},\ \bibinfo {pages} {3379} (\bibinfo {year} {2019})}\BibitemShut {NoStop}%
\bibitem [{\citenamefont {Yin}(2025)}]{Yin_2025}%
  \BibitemOpen
  \bibfield  {author} {\bibinfo {author} {\bibfnamefont {Y.}~\bibnamefont {Yin}},\ }\bibfield  {title} {\bibinfo {title} {Localization-delocalization transition in noninteger-charged electron wave packets},\ }\href {https://doi.org/10.1103/physrevb.111.075402} {\bibfield  {journal} {\bibinfo  {journal} {Physical Review B}\ }\textbf {\bibinfo {volume} {111}},\ \bibinfo {pages} {075402} (\bibinfo {year} {2025})}\BibitemShut {NoStop}%
\bibitem [{\citenamefont {Yue}\ and\ \citenamefont {Yin}(2019)}]{Yue_2019}%
  \BibitemOpen
  \bibfield  {author} {\bibinfo {author} {\bibfnamefont {X.~K.}\ \bibnamefont {Yue}}\ and\ \bibinfo {author} {\bibfnamefont {Y.}~\bibnamefont {Yin}},\ }\bibfield  {title} {\bibinfo {title} {Normal and anomalous electron-hole pairs in a quantum conductor driven by a voltage pulse},\ }\href {https://doi.org/10.1103/physrevb.99.235431} {\bibfield  {journal} {\bibinfo  {journal} {Physical Review B}\ }\textbf {\bibinfo {volume} {99}},\ \bibinfo {pages} {235431} (\bibinfo {year} {2019})}\BibitemShut {NoStop}%
\bibitem [{\citenamefont {Lepage}(1997)}]{Lepage_1997}%
  \BibitemOpen
  \bibfield  {author} {\bibinfo {author} {\bibfnamefont {G.~P.}\ \bibnamefont {Lepage}},\ }\bibfield  {title} {\bibinfo {title} {{How to renormalize the Schrodinger equation}},\ }in\ \href@noop {} {\emph {\bibinfo {booktitle} {{8th Jorge Andre Swieca Summer School on Nuclear Physics}}}}\ (\bibinfo {year} {1997})\ p.\ \bibinfo {pages} {135}\BibitemShut {NoStop}%
\bibitem [{\citenamefont {Delamotte}(2004)}]{Delamotte_2004}%
  \BibitemOpen
  \bibfield  {author} {\bibinfo {author} {\bibfnamefont {B.}~\bibnamefont {Delamotte}},\ }\bibfield  {title} {\bibinfo {title} {A hint of renormalization},\ }\href {https://doi.org/10.1119/1.1624112} {\bibfield  {journal} {\bibinfo  {journal} {American Journal of Physics}\ }\textbf {\bibinfo {volume} {72}},\ \bibinfo {pages} {170} (\bibinfo {year} {2004})}\BibitemShut {NoStop}%
\bibitem [{Note1()}]{Note1}%
  \BibitemOpen
  \bibinfo {note} {This region is smaller than the region $\varphi \in [16, 512]$, which we have used in Ref.~\cite {Yin_2025}. This has little effect on the scaling function and correlation length, but can give a slightly better quality of the data collapse.}\BibitemShut {Stop}%
\bibitem [{\citenamefont {W.}\ \emph {et~al.}(1965)\citenamefont {W.}, \citenamefont {Abramowitz},\ and\ \citenamefont {Stegun}}]{Abramowitz_1965}%
  \BibitemOpen
  \bibfield  {author} {\bibinfo {author} {\bibfnamefont {J.~W.}\ \bibnamefont {W.}}, \bibinfo {author} {\bibfnamefont {M.}~\bibnamefont {Abramowitz}},\ and\ \bibinfo {author} {\bibfnamefont {I.~A.}\ \bibnamefont {Stegun}},\ }\href@noop {} {\emph {\bibinfo {title} {Handbook of Mathematical Functions with Formulas, Graphs, and Mathematical Tables}}},\ Vol.~\bibinfo {volume} {19}\ (\bibinfo  {publisher} {JSTOR},\ \bibinfo {year} {1965})\ p.\ \bibinfo {pages} {147}\BibitemShut {NoStop}%
\bibitem [{\citenamefont {Moskalets}(2016)}]{moskalets_2016_fract}%
  \BibitemOpen
  \bibfield  {author} {\bibinfo {author} {\bibfnamefont {M.}~\bibnamefont {Moskalets}},\ }\bibfield  {title} {\bibinfo {title} {Fractionally charged zero-energy single-particle excitations in a driven fermi sea},\ }\href {https://doi.org/10.1103/physrevlett.117.046801} {\bibfield  {journal} {\bibinfo  {journal} {Physical Review Letters}\ }\textbf {\bibinfo {volume} {117}},\ \bibinfo {pages} {046801} (\bibinfo {year} {2016})}\BibitemShut {NoStop}%
\end{thebibliography}%
\end{document}